\documentclass[journal]{IEEEtran}

\usepackage{amsmath}
\usepackage{mathrsfs}
\usepackage{pifont}
\usepackage{bbding}
\usepackage{amsmath,epsfig}
\usepackage{stmaryrd}
\usepackage{amssymb}
\usepackage{amsfonts}
\usepackage{epic}
\usepackage{graphicx}
\usepackage{curves}
\usepackage{cite}

\usepackage{algorithm}

\usepackage{algpseudocode}
\usepackage{cases}
\usepackage{stfloats}
\usepackage{latexsym}
\usepackage{epstopdf}
\usepackage{epic}
\usepackage{bm}
\usepackage{multirow}
\usepackage{xcolor}
\usepackage{mathrsfs}
\usepackage{pifont}
\usepackage{bbding}
\usepackage{amsmath,epsfig}
\usepackage{mathbbold}
\usepackage{stmaryrd}
\usepackage{amssymb}
\usepackage{amsfonts}
\usepackage{epic}
\usepackage{graphicx}
\usepackage{curves}
\usepackage{algorithm}
\usepackage{algpseudocode}

\usepackage{stfloats}
\usepackage{latexsym}
\usepackage{epstopdf}
\usepackage{epic}
\usepackage{multirow}

\usepackage{subfigure}
\usepackage{color}

\usepackage{amsmath}
\usepackage{mathrsfs}
\usepackage{pifont}
\usepackage{bbding}
\usepackage{amsmath,epsfig}
\usepackage{mathbbold}
\usepackage{stmaryrd}
\usepackage{amssymb}
\usepackage{amsfonts}
\usepackage{epic}
\usepackage{graphicx}
\usepackage{curves}
\usepackage{cite}

\usepackage{algorithm}

\usepackage{algpseudocode}

\usepackage{algpseudocode}

\usepackage{stfloats}
\usepackage{latexsym}
\usepackage{epstopdf}
\usepackage{epic}
\usepackage{bm}
\usepackage{multirow}
\usepackage{xcolor}
\usepackage{subfigure}

\newcommand{\q}{\mathbf q}

\newtheorem{lemma}{Lemma}

\begin{document}
\title{Joint Computation and Communication Design for UAV-Assisted Mobile Edge Computing in IoT}
\author{
\IEEEauthorblockN{ Tiankui~Zhang,~Yu~Xu, Jonathan Loo, Dingcheng~Yang,~Lin~Xiao }

\thanks{
This work was supported by National Natural Science Foundation of China under Grants 61971060 and 61703197.
(Co-corresponding author: Lin Xiao)}
\thanks{Tiankui Zhang, Yu~Xu are with
Beijing University of Posts and Telecommunications, Beijing 100876, China (e-mail: \{zhangtiankui, xuyu56\}@bupt.edu.cn);}
\thanks{Jonathan Loo is with the School of Computing and Engineering, University of West London,  London W5 5RF, U.K. (e-mail: jonathan.loo@uwl.ac.uk);}
\thanks{Dingcheng~Yang and Lin~Xiao are with the Information Engineering School, Nanchang University, Nanchang 330031, China (e-mail: \{yangdingcheng, xiaolin\}@ncu.edu.cn).}
}

\maketitle

\begin{abstract}
Unmanned aerial vehicle (UAV)-assisted mobile edge computing (MEC) system is a prominent concept, where a UAV equipped with a MEC server is deployed to serve a number of terminal devices (TDs) of Internet of Things (IoT) in a finite period. In this paper, each TD has a certain latency-critical computation task in each time slot to complete. Three computation strategies can be available to each TD. First, each TD can operate local computing by itself. Second, each TD can  partially offload task bits to the UAV for computing. Third, each TD can choose to offload task bits to access point (AP) via UAV relaying. We propose a new optimization problem formulation that aims to minimize the total energy consumption including communication-related energy, computation-related energy and UAV's flight energy by optimizing the bits allocation, time slot scheduling and power allocation as well as UAV trajectory design. As the formulated problem is  non-convex and difficult to find the optimal solution, we solve the problem by two parts, and obtain the near optimal solution with within a dozen of iterations. Finally, numerical results are given to validate the proposed algorithm, which is verified to be efficient and superior to the other benchmark cases.
\end{abstract}

\begin{IEEEkeywords}
Internet of Things, mobile edge computing,  resource allocation, trajectory optimization, UAV communication.
\end{IEEEkeywords}
\section{Introduction}

Recently, with the advancement in Internet of Things (IoT) technology, various up-to-date applications, e.g., the augmented reality (AR), virtual reality (VR), autonomous driving and agriculture monitoring, are changing our experience. Some  terminal  devices (TDs) related to the Internet of Things (IoT) such as smart phones, monitoring sensors and wearable devices spring up in our life \cite{R1}\cite{R2}. However, the computation demands for IoT devices are also becoming higher while the computing capacity of these devices is limited. Mobile edge computing (MEC) is considered as a new technology to overcome the limitations by providing cloud-like computing. By deploying computing resource in close proximity to IoT devices (i.e, locating MEC servers at a wireless access point (AP) or base station), it can efficiently reduce the delay and save the computation resource at these devices by the way of computation task offloading \cite{R3}\cite{R4}. Therefore, MEC has the potential to provide  the service of solving the computation-intensive  and latency-critical tasks for devices.  In general, the MEC server deployment is fixed, which means that it can not exploit its mobility to move closer to    TDs, by which the latency or energy consumption of the devices would be further reduced.

Due to the high flexible mobility,  unmanned aerial vehicle (UAV) has attracted significant research interest in academia \cite{R8}-\cite{add2}. In wireless communications, UAV has been applied in various scenarios, such as nonorthogonal multiple access (NOMA) networks \cite{add3}, mmWave communications \cite{add4} and caching \cite{add5} \cite{add6}. Also, the three-dimensional coverage performance for cellular network-connected UAVs that act as aerial users is also investigated in \cite{add7}. In addition, UAV relaying \cite{R12}-\cite{R14} is also an important application that can efficiently expand the communication coverage. By utilizing UAV as a relay, two users with communication channel blocked can be linked. This gives a new method to help local resource-limited users access to the remote resources.

The new setup by utilizing UAV to assist computing in MEC systems poses new opportunities to solve the challenges in communication and computation design, and several  prior related works have been done for this \cite{UAV_MEC_1}-\cite{UAV_MEC_11}. Specifically, the work \cite{UAV_MEC_1} considers that a UAV is deployed to provide computation service for TDs, and a minimization problem of sum of the maximum delay among users is proposed by optimizing the offloading ratio, users scheduling and UAV trajectory. In work \cite{UAV_MEC_2}, the computation rate maximization problem in a UAV-assisted MEC is investigated. The authors in \cite{UAV_MEC_3} focus on minimizing the average weighted energy consumption of TDs, and the optimal solution is obtained by decomposing the primal problem into three subproblems. The authors in \cite{UAV_MEC_4} investigate computation energy consumption of mobile terminal minimization problem, but the UAV trajectory is not optimized. Hua \emph{et al.} \cite{UAV_MEC_6} consider a UAV to help TD offload bits, the TD can compute locally as well as can offload bits to the UAV. Besides, the works \cite{UAV_MEC_7} and \cite{UAV_MEC_8} study the UAV energy minimization problem and task completion time minimization problem  in cellular-connected UAV MEC systems, respectively. Bai  \emph{et al.} \cite{UAV_MEC_9} focus on the security in UAV-assisted MEC systems, where a potential passive eavesdropper can capture the offloading bits from the UAV to AP via eavesdropping channel. Also, Du \emph{et al.} \cite{UAV_MEC_10} study the energy efficiency of the UAV in a MEC system, by minimizing the hovering energy and computation energy of the UAV. In addition, the work \cite{UAV_MEC_11} study the problem described as the offloading bits from users to UAV maximization, subject to each user's quality of service (QoS).  These existing works related to UAV-assisted MEC systems mainly focus on the computing bits offloading only between UAV(s) and users.

Different from the existing works, we propose a framework in which the UAV acts as a relay to assist bits offloading for TDs. Specifically, the UAV can not only  provide the computation service but also can provide the communication service for TDs by forwarding the received bits to AP for remote computation. Thus, our proposed framework further enhances the computing ability of the MEC systems.

The UAV-assisted MEC systems in IoT are studied in this paper, in which the UAV is considered as a helper that  not only helps computing the bits offloaded from TDs but also acts as a decode-and-forward (DF)   relay to assist task bits transmit from TDs to AP. Considering the practical terrible channel environment between the TDs and remote AP, and in order to clearly shed light on the essence of our proposed system, it is assumed that the direct communication links between TDs and AP are blocked. Also, the total energy on UAV  is enough to support propulsion and complete the task during the period. These TDs need to process their collected data, such as the video file, temperature information and movement data, they need to transmit a part of task bits to the UAV for processing if they are unable to compute locally. For a given period, each TD needs to complete the required  latency-critical task in per time slot. In addition, considering the AP is located on the ground, it can be equipped with a or several powerful MEC server(s). Thus, the maximum computing rate at the AP would be much larger than the bits offloading rate from the UAV in our setup. Therefore, it is reasonable to assume that the computing time at AP in each time slot is neglectable. Our goal is to minimize the sum energy of communication-related energy, computation-related energy and UAV's flight energy subject to the constraints on communication and computation resource allocation, computation causality constraint and UAV trajectory design. In our design, the UAV's mobility is restricted by the maximum speed and initial/final location, and it serves the TDs in an orthogonal frequency-division multiple access (OFDMA) manner. In summary, the main contributions of this work are  presented as follows.
\begin{itemize}
  \item We propose a new framework of UAV-assisted MEC system in IoT.  Our proposed framework fills the gap that jointly considers the task offloading strategy and UAV relay communication in MEC systems, which provides useful insights and guidelines for designing the similar problems in practice.  In our design, the required computation bits can be computed by TDs locally, or offloaded to the UAV for computing. Besides, the required task bits also can be transmitted to the AP for computing via UAV relaying. This mode can further expand the computation resource scale and provide a new opportunity to solve the challenges in traditional MEC systems.
  \item We formulate a total energy consumption minimization problem, by optimizing the computation bits allocation, time slot scheduling, transmit power allocation and UAV trajectory. A problem decomposition method is adopted to tackle the non-convex  problem in two parts that are solved by the Lagrangian duality method and successive convex approximation (SCA) technique, respectively.
  \item We present the numerical results that show the superiorities of our proposed design, as compared with other benchmark designs. On the one hand, the proposed algorithm can be guaranteed to converge within a dozen of iterations. On the other hand, the total energy consumption obtained by the proposed algorithm is always lowest, indicating the significant effectiveness of our design.
\end{itemize}


\section{System Model and Problem Formulation}
 \begin{figure}
\setlength{\abovecaptionskip}{0.cm}
\centering
\includegraphics[width=0.4\textwidth]{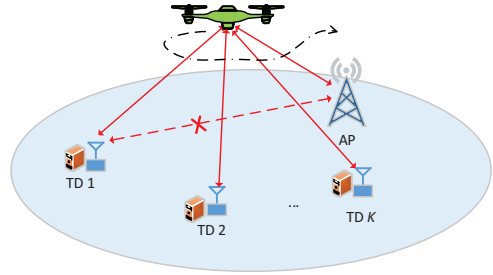}
\caption{The illustration of a UAV-enabled MEC system.}\label{system_model}
\setlength{\belowcaptionskip}{-0.2cm}
\vspace{-1em}
\end{figure}

Consider a UAV-assisted MEC system in IoT as shown in Fig. \ref{system_model}, where a UAV is deployed as a mobile DF relaying over an area of interest. The UAV is dispatched to help computation bits of TDs that are denoted by a set $\mathcal{K}=\{1,2,...,K\}$ transmit to AP equipped with MEC functionality for computing.  For convenience, we use the notation $u_k$ to denote TD $k$ in this paper. Meanwhile, the UAV is also equipped with a MEC server to provide computation operation for TDs. The UAV, each TD and AP are assumed to be equipped with one single antenna, respectively. Without loss any generality, we assume that the UAV flies from an initial location $\mathbf{q}_0$  to final location  $\mathbf{q}_F$. The flight altitude is fixed at $H$ that effectively avoids any collisions. The period time for the UAV flight is expressed by $T$. Considering a 3D Cartesian coordinate system, the UAV's location projected on the horizontal plane in any time instant $t\in [0,T]$ can be represented by $\mathbf{q}(t)=\{x(t),y(t)\}$. In addition, the locations of AP and each TD $k\in\mathcal{K}$ are fixed at $\mathbf{w_a}=\left(x_a,y_a\right)$ and $\mathbf{w_k}=\left(x_k,y_k\right)$, respectively. For convenience, we use sufficiently small constant $\delta_t$ to divide the period $T$ into $N$ slots with equal size, which are expressed by a set $\mathcal{N}=\{1,2,...,N\}$. In each time slot, the UAV can be considered to be static.  Thus, the UAV's location in any time slot $n\in \mathcal{N}$ can be denoted by $\mathbf{q}[n]=\{x[n],y[n]\}$, with $\mathbf{q}(t)=\mathbf{q}(\delta_tn)=\mathbf{q}[n]$. Hence, the distance between the TD $k$ and UAV/helper in each time slot $n\in \mathcal{N}$ can be denoted by $d_{u_kh}[n]=\sqrt{H^2+||\mathbf q[n]-\mathbf{w_k}||^2}$, where $||\cdot||$ denotes Euclidean norm. Similarly, the distance between the UAV and AP in each time slot can be denoted by $d_{ha}[n]=\sqrt{H^2+||\mathbf q[n]-\mathbf{w_a}||^2}$.

For each TD $k\in\mathcal{K}$, it has a  latency-critical computation task requirement in each time slot $n\in \mathcal{N}$, i.e., each user needs to complete at least $L_{k,n}^{\min}$ bits of computation task in each time slot $n$. Considering its limited computing ability, each TD can offload the computation bits to the UAV via wireless transmit for either computing or relaying. Let $l_{u,k}[n]$, $l_{h,k}[n]$ and $l_{a,k}[n]$ denote the amount of computation bits allocated for local computing, offloading to UAV for computing and offloading to AP for computing via relaying (or offloading to UAV for relaying) in each time slot, respectively. Thus we have,
\begin{align}\label{computation_variable}
l_{u,k}[n]\geq0,l_{h,k}[n]\geq0,l_{a,k}[n]\geq0,
\end{align}
\begin{align}\label{computation_variable2}
l_{u,k}[n]+l_{h,k}[n]+l_{a,k}[n]\geq L_{k,n}^{\min}, \forall k, n.
\end{align}

\subsection{Communication Model}

\begin{figure}
\setlength{\abovecaptionskip}{0.cm}
\centering
\includegraphics[width=0.5\textwidth]{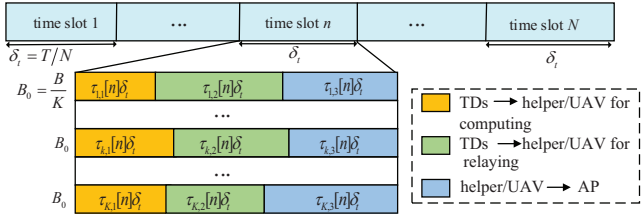}
\caption{The illustration of computation bits offloading protocol.}\label{protocol}
\vspace{-0.8cm}
\end{figure}

Note that the delay and energy consumption for results sending back from UAV to TDs and that from AP to UAV are omitted since the size of results is much smaller than offloaded data size \cite{UAV_MEC_2}\cite{UAV_MEC_8}.  As shown in Fig. \ref{protocol}, we consider a computation bits offloading protocol of each TD. Specifically, in each time slot $n\in \mathcal{N}$, the TDs can offload their tasks  to the UAV for computing. It is assumed that the wireless channel between the UAV and TD $k$ is dominated by LoS link \cite{UAV_MEC_2}\cite{R27}, hence the channel between the UAV and TDs and that between the UAV and AP are modeled by the free space path loss model. Thus, the channel power gain from TD $k$ to UAV is given as
  \begin{align}
h_{u_kh}[n]=\beta_0d^{-2}_{u_kh}[n]=\frac{\beta_0}{H^2+||\q[n]-\mathbf{w_k}||^2},
\end{align}
where $\beta_0$  denotes the channel gain at the reference distance $d_0=1$ meter. Besides, the TDs also can offload their tasks to AP via UAV. Thus, the channel power gain from the UAV to AP is obtained as
  \begin{align}
h_{ha}[n]=\beta_0d^{-2}_{ha}[n]=\frac{\beta_0}{H^2+||\q[n]-\mathbf{w_a}||^2}.
\end{align}

In Fig. \ref{protocol}, for each TD $k$, each time slot $n\in\mathcal{N}$ is divided into 3 subslots that are dedicated to be allocated for bits offloading to UAV for computing, bits offloading to UAV for relaying and bits forwarding to AP from UAV, respectively. The size of each subslot is determined by variable $\tau_{k,m}[n]$, with $m=\{1,2,3\}$, which satisfies the following constraints,
\begin{align}\label{tau_constraint}
\sum\limits_{m=1}^{3}\tau_{k,m}[n]\leq1,\forall k, n,
\end{align}
\begin{align}\label{tau_constraint2}
0\leq\tau_{k,m}[n]\leq1,\forall k, n,m=\{1,2,3\}.
\end{align}

 It is assumed that an OFDMA is applied in the system. The total available bandwidth $B$ is equally divided into $K$ sub-carriers with size of $B_0=\frac{B}{K}$ for each TD. The transmit power of TD $k$ for offloading bits to UAV for computing in each time slot $n$ is denoted by $p_{k,1}[n]$. Therefore, the achievable offloading rate in bits-per-second (bps) from the TD $k$ to UAV for computing is denoted as
\begin{align}\label{R_U2H1}
r_{uh}&\left( p_{k,1}[n],\mathbf{q}[n] \right) =B_0\log_2\left( 1+\frac{p_{k,1}[n]h_{u_kh}[n]}{N_0B_0} \right) \nonumber \\
&=B_0\log_2\left( 1+\frac{p_{k,1}[n]\gamma_0}{||\mathbf{q}[n]-\mathbf{w}_k[n]||^2+H^2} \right),\forall k, n,
\end{align}
where $\gamma_0=\frac{\beta_0}{N_0B_0}$ denotes the reference received signal-to-noise ratio (SNR) at UAV for $d_0=1$ meter, and $N_0$ denotes noise power dense at the UAV. Assume that the transmit power of TD $k$ to UAV for relaying in the second subslot with duration of $\delta_t\tau_{k,2}[n]$ is denoted by $p_{k,2}[n]$. Thus, the achievable offloading rate in bps from the TD $k$ to UAV for relaying is given as
\begin{align}\label{R_U2H2}
r_{uh}&\left( p_{k,2}[n],\mathbf{q}[n] \right) =B_0\log_2\left( 1+\frac{p_{k,2}[n]h_{u_kh}[n]}{N_0B_0} \right) \nonumber \\
&=B_0\log_2\left( 1+\frac{p_{k,2}[n]\gamma_0}{||\mathbf{q}[n]-\mathbf{w}_k[n]||^2+H^2} \right),\forall k, n.
\end{align}

Similarly, the achievable forwarding rate from the UAV to AP in bps is given as
\begin{align}\label{R_H2A}
r_{ha}&\left( p_{k,3}[n],\mathbf{q}[n] \right) =B_0\log_2\left( 1+\frac{p_{k,3}[n]h_{ha}[n]}{N_1B_0} \right) \nonumber \\
&=B_0\log_2\left( 1+\frac{p_{k,3}[n]\gamma_1}{||\mathbf{q}[n]-\mathbf{w}_a[n]||^2+H^2} \right),\forall k, n,
\end{align}
 where $\gamma_1=\frac{\beta_0}{N_1B_0}$ denotes the reference received SNR at AP for $d_0=1$ meter, and  $N_1$ denotes noise power dense at AP.

  In addition, we assume that the UAV is able to store the  unprocessed offloading bits from TDs in its memory if the offloading rate exceeds its computing ability.  Consequently,  we can obtain the following computation causality condition,
 \begin{align}\label{lh_computation}
\sum\limits_{i=1}^{n}l_{h,k}[i]\leq\sum\limits_{i=1}^{n}\delta_t\tau_{k,1}[n]r_{uh}\left( p_{k,1}[n],\mathbf{q}[n]\right),\forall k, n,
\end{align}

Assuming that the processing delay at the DF relay is one subslot, the computing bits $l_{a,k}[n]$ should satisfy the expression shown in \eqref{la_computation}.

\newcounter{mytempeqncnt}
\begin{figure*}[!t]
\begin{align}\label{la_computation}
l_{a,k}[n] \leq \min(\delta_t\tau_{k,2}[n]r_{uh}\left( p_{k,2}[n],\mathbf{q}[n]\right),\delta_t\tau_{k,3}[n]r_{ha}\left( p_{k,3}[n],\mathbf{q}[n]\right)),\forall k, n,
\end{align}
\hrulefill
\end{figure*}

 In this model, the total communication-related energy consumption is considered, given by
 \begin{align}\label{communication energy}
&E_{comm}=\delta_t\sum\limits_{m=1}^{3}\sum\limits_{n=1}^{N}\sum\limits_{k=1}^{K}\left(\tau_{k,m}[n]p_{k,m}[n] \right)
\end{align}

 \subsection{Computation Model}

Let $c_u^k>0$ denotes the required CPU cycles for computing each one bit at the user $k$, and $\kappa_u^k>0$ represents the effective capacitance coefficient effected by chip architecture at TD $k$ \cite{R28_b}. It is assumed that all TDs have same CPU cycles and capacitance coefficient, i.e., $c_u^k=c_u$, $\kappa_u^k=\kappa_u, \forall k$.  In order to help TDs complete computation tasks in each time slot, as shown in \eqref{computation_variable2}, we assume that the CPU cycles and capacitance coefficient of the MEC server at the UAV are $c_h>0$ and $\kappa_h>0$, respectively. In addition, the computation capacity of the AP is assumed to be sufficiently powerful so that the computing time at the AP can be negligible in our setup. The maximum CPU frequency at each TD and UAV is denoted by $f_u^{max}$ and $f_h^{max}$, respectively. As a result, in any time slot, we have
\begin{align}\label{lu_constraint}
c_ul_{u,k}[n]\leq \delta_tf_{u}^{max}, \forall k, n,
\end{align}
\begin{align}\label{lh_constraint}
c_hl_{h,k}[n]\leq \delta_t\bar{f}_{h}^{max}, \forall k, n,
\end{align}
where $\bar{f}_{h}^{max}=\frac{f_{h}^{max}}{K}$ indicates that the total frequency of the UAV is equally divided into $K$ parts that are allocated to each TD, respectively. Based on \cite{UAV_MEC_8}, the energy consumption in each time slot for local computing is expressed as
\begin{align}\label{E_u}
E_{comp}^{u,k}[n]= \frac{k_u\left(c_ul_{u,k}[n]\right)^3}{\delta_t^2}, \forall k, n.
\end{align}

Similarly, the energy consumption in each time slot for UAV computing is expressed as
\begin{align}\label{E_h}
E_{comp}^{h}[n]= \sum\limits_{k=1}^{K}\left( \frac{k_h\left(c_hl_{h,k}[n]\right)^3}{\delta_t^2}   \right), \forall n,
\end{align}
it is worth mentioning that in the first time slot of $n=1$, the available time duration for UAV computing is $\left(\delta_t-\delta_t\tau_{k,1}[1]\right)$ s. However, considering the time slot size $\delta_t$ in our design is chosen to be quite small so that we have $\delta_t\tau_{k,1}[1]\ll T$. Thus, the computing time  for the UAV in first time slot $n=1$ can be approximated to be $\delta_t$. As a result, the total computation-related energy consumption can be denoted as
\begin{align}\label{E_comp}
E_{comp}=& \sum\limits_{n=1}^{N}\sum\limits_{k=1}^{K} E_{comp}^{u,k}[n]  + \sum\limits_{n=1}^{N} E_{comp}^{h}[n].
\end{align}
\vspace{-0.5em}
 \subsection{UAV Mobility and Flight Energy Consumption Model}
In the proposed system, an altitude-fixed rotary-wing UAV is considered. In practice, this UAV  flies from an initial location to a final location, during which its speed is constrained by a maximum speed $V_{\max}$. Hence, we have
\begin{subequations}
\begin{align}\label{T1}
\mathbf{q}[1]=\mathbf{q}_0,
\end{align}
\begin{align}\label{T2}
\mathbf{q}[N+1]=\mathbf{q}_F,
\end{align}
\begin{align}\label{T3}
||\mathbf{q}[n+1]-\mathbf{q}[n]||^2\leq\left(\delta_tV_{\max}\right)^2,\forall n.
\end{align}
\end{subequations}

Based on \cite{E_model}, the power consumption of flight for rotary-wing UAV is modeled as
\begin{align}
P&\left(||\mathbf{v}[n]||\right)=P_0\left(1+\frac{3||\mathbf{v}[n]||^2}{U_{tip}^2}\right)+ \nonumber\\
&P_i\left(\sqrt{1+\frac{||\mathbf{v}[n]||^4}{4v_0^4}}-\frac{||\mathbf{v}[n]||^2}{2v_0^2}\right)^{\frac{1}{2}}+\frac{1}{2}d_0\rho sA||\mathbf{v}[n]||^3 \nonumber\\
&\quad\quad\quad\quad\quad\quad\quad\quad\quad\quad\quad\quad\quad\quad\quad\quad ,\forall n\in\mathcal{N}, \label{p_consumption}
\end{align}
where $P_0$ and $P_i$  represent the blade profile power and induced power in hovering status, respectively. The other parameters of $U_{tip}$, $v_0$, $d_0$, $\rho$, $s$ and $A$ related to the UAV's aerodynamics are given in Table I in Section IV based on the work \cite{E_model}. To achieve the UAV's peed $\mathbf{v}[n]$, we have
\begin{align}
\mathbf{v}[n]=\frac{\mathbf{q}[n+1]-\mathbf{q}[n]}{\delta_t},\forall n\in\mathcal{N}. \label{obtain_v}
\end{align}

Thus, the UAV's flight energy consumption during the period time is expressed as
\begin{align}
E_{fly}=\delta_t\sum\limits_{n=1}^{N}P\left(||\mathbf{v}[n]||\right), \forall n\in\mathcal{N}. \label{E_consumption}
\end{align}

 \subsection{Problem Formulation}
According to the discussion above, we formulate the objective problem as a sum of communication-related energy consumption and computation-related energy consumption minimization, which is subjected to task allocation, time slot scheduling, transmit power allocation and UAV trajectory design. Specifically, the problem is formulated as
\begin{subequations}
\begin{align}
 \text{(P1):}&\min \limits_{\mathbf{L},\boldsymbol{\tau},\mathbf{P},\mathbf{Q}} E_{comm} +  E_{comp} + wE_{fly}\nonumber\\
 \text{s.t.} ~~& \eqref{computation_variable},\eqref{computation_variable2},\eqref{tau_constraint},\eqref{tau_constraint2},\eqref{lh_computation},\eqref{la_computation},\eqref{lu_constraint},\eqref{lh_constraint},\eqref{T1}\text{-}\eqref{T3}
 ,\eqref{obtain_v} \nonumber\\
&0\leq p_{k,1}[n]\leq P_{u}^{max},\forall k, n,\\
&0\leq p_{k,2}[n]\leq P_{u}^{max},\forall k, n, \\
&0\leq p_{k,3}[n]\leq P_h^{max},\forall k, n.
\end{align}
\end{subequations}
where  $\mathbf{L}=\left\{l_{u,k}[n],l_{h,k}[n],l_{a,k}[n] \right\}$, $\boldsymbol{\tau}=\left\{\tau_{k,m}[n]\right\}_{m=1}^3$, $\mathbf{P}=\left\{p_{k,i}[n]\right\}_{i=1}^3$, $\mathbf{Q}=\left\{ \mathbf{q}[n],\mathbf{v}[n] \right\}_{n=1}^N$, $P_{u}^{max}$ and $P_{h}^{max}$ stand for the maximum transmit power at each TD and UAV, respectively. Like \cite{UAV_MEC_2}, let $w$ denote the given weight with regard to the UAV's flight energy consumption to ensure the fairness for TDs.

Obviously, problem (P1) is a non-convex problem due to the non-convexity in the constraints \eqref{lh_computation} and \eqref{la_computation} as well as in the objective function. To tackle this, the primal problem (P1) is decomposed into two manageable subproblems, which are analyzed in the following sections.
\vspace{-0.5em}
\section{Energy Minimization with Fixed Trajectory}
For any given UAV trajectory $\mathbf{Q}$, and let $E_{k,m}[n]=t_{k,m}[n]p_{k,m}[n]$ and  $t_{k,m}[n]=\delta_t\tau_{k,m}[n]$, with $m=\{1,2,3\}$. The primal problem (P1) is formulated as
problem (P2),
\begin{small}
\begin{subequations}
\begin{align}
 \text{(P2):}\min \limits_{\mathbf{L},\boldsymbol{t},\mathbf{E}} \sum\limits_{n=1}^{N}\sum\limits_{k=1}^{K} &(E_{k,1}[n]+E_{k,2}[n]+E_{k,3}[n] \nonumber\\
 &~~~~~~~+ \frac{k_u\left(c_ul_{u,k}[n]\right)^3}{\delta_t^2} + \frac{k_h\left(c_hl_{h,k}[n]\right)^3}{\delta_t^2} )   \nonumber
 \end{align}
 \begin{align}
\text{s.t.} ~~& \sum\limits_{i=1}^{n}l_{h,k}[i]\leq\sum\limits_{i=1}^{n}t_{k,1}[i]r_{uh}\left( \frac{E_{k,1}[i]}{t_{k,1}[i]} \right),\forall k, n, \label{C1}\\
&l_{a,k}[n] \leq t_{k,2}[n]r_{uh}\left( \frac{E_{k,2}[n]}{t_{k,2}[n]} \right), \forall k, n,\label{C2}\\
 &l_{a,k}[n]\leq t_{k,3}[n]r_{ha}\left( \frac{E_{k,3}[n]}{t_{k,3}[n]} \right), \forall k, n, \label{C3}\\
 &l_{u,k}[n]+l_{h,k}[n]+l_{a,k}[n]\geq L_{k,n}^{min}, \forall k, n,\label{C4}\\
 &c_ul_{u,k}[n]\leq \delta_tf_{u}^{max}, \forall k, n,\label{C5}\\
&c_hl_{h,k}[n]\leq \delta_t\bar{f}_{h}^{max}, \forall k, n, \label{C6}\\
&0\leq E_{k,1}[n]\leq t_{k,1}[n]P_{u}^{max},\forall k, n,\label{C7}\\
&0\leq E_{k,2}[n]\leq t_{k,2}[n]P_{u}^{max},\forall k, n, \label{C8}\\
&0\leq E_{k,3}[n]\leq t_{k,3}[n]P_h^{max},\forall k, n ,\label{C9}\\
&l_{u,k}[n]\geq 0, l_{h,k}[n]\geq 0, l_{a,k}[n]\geq 0,\forall k, n, \label{C10}\\
&\sum\limits_{m=1}^{3}t_{k,m}[n]\leq\delta_t,\forall k, n,m=\{1,2,3\} \label{C11}\\
&0\leq t_{k,m}[n]\leq \delta_t,\forall k, n,m=\{1,2,3\},\label{C12}
\end{align}
\end{subequations}
\end{small}
where $\mathbf{E}=\{E_{k,m}[n]\}_{m=1}^3$ and $\mathbf{t}=\left\{t_{k,m}[n]\right\}_{m=1}^3$.

\begin{lemma}\label{lemma1}
Problem (P2) is a convex problem.
\end{lemma}
\begin{IEEEproof}
Firstly, the objective function of problem (P2) is convex with respect to $\mathbf{E}$, $l_{u,k}[n]$ and $l_{h,k}[n]$. Then, it can be easy find that the expressions in constraints \eqref{C4}-\eqref{C12} are linear. $f(x,t)=t\log(1+\frac{x}{t})$ with $t>0$, is concave \cite{Boyd}. Therefore, the expressions $t_{k,1}[i]r_{uh}\left( \frac{E_{k,1}[i]}{t_{k,1}[i]} \right)$, $t_{k,2}[n]r_{uh}\left( \frac{E_{k,2}[n]}{t_{k,2}[n]} \right)$ and $t_{k,3}[n]r_{ha}\left( \frac{E_{k,3}[n]}{t_{k,3}[n]} \right)$ respectively in constraints \eqref{C1}-\eqref{C3} are concave. Thus, problem (P2) is proofed to be convex.
\end{IEEEproof}

In order to achieve the closed-form solutions and give more insights into the proposed problem (P2), we choose the Lagrange duality method to solve this problem in this paper.
By introducing the nonnegative dual variables $\lambda_{k,n}$, $\mu_{k,n}$, $\nu_{k,n}$, $\omega_{k,n}$ and $\eta_{k,n}$ that are corresponding to the constraints \eqref{C1}-\eqref{C4} and \eqref{C11}, respectively, and let $\boldsymbol{\lambda}=\{\lambda_{k,n}\}$, $\boldsymbol{\mu}=\{\mu_{k,n}\}$, $\boldsymbol{\nu}=\{\nu_{k,n}\}$, $\boldsymbol{\omega}=\{\omega_{k,n}\}$ and $\boldsymbol{\eta}=\{\eta_{k,n}\}$, then the Lagrange function of problem (P2) is
\begin{small}
\begin{align}\label{L_F}
&\mathcal{L}\left( \mathbf{L},\mathbf{t},\mathbf{E},\boldsymbol{\lambda}, \boldsymbol{\mu},\boldsymbol{\nu},\boldsymbol{\omega},\boldsymbol{\eta} \right)=\sum\limits_{n=1}^{N}\sum\limits_{k=1}^{K} E_{k,1}[n]+\sum\limits_{n=1}^{N}\sum\limits_{k=1}^{K}E_{k,2}[n]\nonumber\\
 &+\sum\limits_{n=1}^{N}\sum\limits_{k=1}^{K}E_{k,3}[n]+ \sum\limits_{n=1}^{N}\sum\limits_{k=1}^{K}\frac{k_u\left(c_ul_{u,k}[n]\right)^3}{\delta_t^2}\nonumber\\
  &+ \sum\limits_{n=1}^{N}\sum\limits_{k=1}^{K}\frac{k_h\left(c_hl_{h,k}[n]\right)^3}{ \delta_t^2}+\sum\limits_{n=1}^{N}\sum\limits_{k=1}^{K}\left( \hat{\lambda}_{k,n}-\omega_{k,n}\right)l_{h,k}[n]\nonumber\\
  &-\sum\limits_{n=1}^{N}\sum\limits_{k=1}^{K}\hat{\lambda}_{k,n}t_{k,1}[n]r_{uh}\left( \frac{E_{k,1}[n]}{t_{k,1}[n]} \right)+\sum\limits_{n=1}^{N}\sum\limits_{k=1}^{K}\eta_{k,n}t_{k,1}[n]\nonumber\\
  &+\sum\limits_{n=1}^{N}\sum\limits_{k=1}^{K}\left(\mu_{k,n}+\nu_{k,n}-\omega_{k,n} \right)l_{a,k}[n]+\sum\limits_{n=1}^{N}\sum\limits_{k=1}^{K}\eta_{k,n}t_{k,2}[n]\nonumber\\
  &-\sum\limits_{n=1}^{N}\sum\limits_{k=1}^{K}\mu_{k,n}t_{k,2}[n]r_{uh}\left( \frac{E_{k,2}[n]}{t_{k,2}[n]} \right)+\sum\limits_{n=1}^{N}\sum\limits_{k=1}^{K}\eta_{k,n}t_{k,3}[n]\nonumber\\
  &-\sum\limits_{n=1}^{N}\sum\limits_{k=1}^{K}\nu_{k,n}t_{k,3}[n]r_{ha}\left( \frac{E_{k,3}[n]}{t_{k,3}[n]} \right)-\sum\limits_{n=1}^{N}\sum\limits_{k=1}^{K}\omega_{k,n}l_{u,k}[n]\nonumber\\
  &+\sum\limits_{n=1}^{N}\sum\limits_{k=1}^{K}\omega_{k,n}L_{k,n}^{min}-\sum\limits_{n=1}^{N}\sum\limits_{k=1}^{K}\eta_{k,n}\delta_t
\end{align}
\end{small}

In \eqref{L_F}, note that $\hat{\lambda}_{k,n}$ is a new defined parameter that satisfies $\hat{\lambda}_{k,n}=\sum\limits_{i=n}^{N}\lambda_{k,i}$. Thus, the dual function of problem (P2) can be denoted by $g\left(\boldsymbol{\lambda}, \boldsymbol{\mu},\boldsymbol{\nu},\boldsymbol{\omega},\boldsymbol{\eta} \right)$, given as
\begin{align}\label{D_F}
g\left(\boldsymbol{\lambda}, \boldsymbol{\mu},\boldsymbol{\nu},\boldsymbol{\omega},\boldsymbol{\eta} \right)=&\min \limits_{\mathbf{L},\boldsymbol{t},\mathbf{E}}\mathcal{L}\left( \mathbf{L},\mathbf{t},\mathbf{E},\boldsymbol{\lambda}, \boldsymbol{\mu},\boldsymbol{\nu},\boldsymbol{\omega},\boldsymbol{\eta} \right) \\
&\text{s.t.} ~~\eqref{C5}\text{-}\eqref{C10}, \eqref{C12}.\nonumber
\end{align}

\begin{lemma}\label{lemma2}
In order to make $g\left(\boldsymbol{\lambda}, \boldsymbol{\mu},\boldsymbol{\nu},\boldsymbol{\omega},\boldsymbol{\eta} \right)$ bounded, the expression of $\left(\mu_{k,n}+\nu_{k,n}-\omega_{k,n} \right)\geq 0$ must hold.
\end{lemma}
\begin{IEEEproof}
Lemma 2 can be shown by contradiction. Assume that $\left(\mu_{k,n}+\nu_{k,n}-\omega_{k,n} \right)< 0$, thus the value of $l_{a,n}[n]$ would $\rightarrow+\infty$ in order to minimize the objective function. Thus, the value of dual function  $g\left(\boldsymbol{\lambda}, \boldsymbol{\mu},\boldsymbol{\nu},\boldsymbol{\omega},\boldsymbol{\eta} \right)$ would be minus infinity. This lemma is proved.
\end{IEEEproof}

As a result, the dual problem of problem (P2) can be written as
\begin{subequations}
\begin{align}
 \text{(D2):}&\max\limits_{\boldsymbol{\lambda}, \boldsymbol{\mu},\boldsymbol{\nu},\boldsymbol{\omega},\boldsymbol{\eta}} g\left(\boldsymbol{\lambda}, \boldsymbol{\mu},\boldsymbol{\nu},\boldsymbol{\omega},\boldsymbol{\eta} \right)\nonumber\\
\text{s.t.} ~~& \boldsymbol{\lambda}\succeq0,\boldsymbol{\mu}\succeq0,\boldsymbol{\nu}\succeq0,\boldsymbol{\omega}\succeq0,\boldsymbol{\eta}\succeq0\\
&\left(\mu_{k,n}+\nu_{k,n}-\omega_{k,n} \right)\geq 0,\forall k,n.
\end{align}
\end{subequations}
Due to problem (P2) is convex, the Slater's condition can be satisfied \cite{Boyd} and thus the strong duality holds between (P2) and (D2). So we can obtain the optimal solution of problem (P2) by solving its dual problem, i.e., problem (D2).

\subsection{Obtaining $g\left(\boldsymbol{\lambda}, \boldsymbol{\mu},\boldsymbol{\nu},\boldsymbol{\omega},\boldsymbol{\eta} \right)$ by Solving Problem \eqref{D_F}}

For any given value of $\left(\boldsymbol{\lambda}, \boldsymbol{\mu},\boldsymbol{\nu},\boldsymbol{\omega},\boldsymbol{\eta} \right)$ in the feasible set of problem (D2), the dual function can be obtained by solving problem \eqref{D_F}. Note the problem \eqref{D_F} can be decomposed into $KN$ independent subproblems, and each one is further decomposed into several subproblems as follows.

\begin{small}
\begin{subequations}
\begin{align}
\text{(L1):}\min \limits_{t_{k,1}[n],E_{k,1}[n]}& E_{k,1}[n] - \hat{\lambda}_{k,n}t_{k,1}[n]r_{uh}\left( \frac{E_{k,1}[n]}{t_{k,1}[n]} \right)\nonumber\\
&+ \eta_{k,n}t_{k,1}[n]\nonumber\\
\text{s.t.} ~~&0\leq E_{k,1}[n]\leq t_{k,1}[n]P_{u}^{max},\forall k, n,\\
&0\leq t_{k,1}[n]\leq\delta_t,\forall k, n.
\end{align}
\end{subequations}

\begin{subequations}
\begin{align}
\text{(L2):}\min \limits_{t_{k,2}[n],E_{k,2}[n]}& E_{k,2}[n] - \mu_{k,n}t_{k,2}[n]r_{uh}\left( \frac{E_{k,2}[n]}{t_{k,2}[n]} \right)\nonumber\\
&+ \eta_{k,n}t_{k,2}[n]\nonumber\\
\text{s.t.} ~~&0\leq E_{k,2}[n]\leq t_{k,2}[n]P_{u}^{max},\\
&0\leq t_{k,2}[n]\leq\delta_t.
\end{align}
\end{subequations}

\begin{subequations}
\begin{align}
\text{(L3):}\min \limits_{t_{k,3}[n],E_{k,3}[n]}& E_{k,3}[n] - \nu_{k,n}t_{k,3}[n]r_{ha}\left( \frac{E_{k,3}[n]}{t_{k,3}[n]} \right)\nonumber\\
&+ \eta_{k,n}t_{k,3}[n]\nonumber\\
\text{s.t.} ~~&0\leq E_{k,3}[n]\leq t_{k,3}[n]P_{h}^{max},\\
&0\leq t_{k,3}[n]\leq\delta_t.
\end{align}
\end{subequations}

\begin{subequations}
\begin{align}
\text{(L4):}\min \limits_{l_{u,k}[n]}& \frac{k_u\left(c_ul_{u,k}[n]\right)^3}{ \delta_t^2}-\omega_{k,n}l_{u,k}[n] \nonumber\\
\text{s.t.} ~~&l_{u,k}[n]\geq0 \\
&c_ul_{u,k}[n]\leq \delta_tf_{u}^{max}.
\end{align}
\end{subequations}

\begin{subequations}
\begin{align}
\text{(L5):}\min \limits_{l_{h,k}[n]}& \frac{k_h\left(c_hl_{h,k}[n]\right)^3}{ \delta_t^2}+\left(\hat{\lambda}_{k,n}-\omega_{k,n}\right)l_{h,k}[n] \nonumber\\
\text{s.t.} ~~&l_{h,k}[n]\geq0 \\
&c_ul_{h,k}[n]\leq \delta_t\bar{f}_{h}^{max}.
\end{align}
\end{subequations}

\begin{align}
\text{(L6):}\min \limits_{l_{a,k}[n]}& \left(\mu_{k,n}+\nu_{k,n}-\omega_{k,n}\right)l_{a,k}[n] \nonumber\\
\text{s.t.} ~~&l_{a,k}[n]\geq0.
\end{align}
\end{small}

For these subproblems, they are all convex  so that their solutions satisfy the Karush-Kuhn-Tucker (KKT) conditions.

\begin{lemma}\label{lemma3}
By solving subproblem (L1) with KKT, the optimal solution can be denoted as

\begin{small}
\begin{subequations}
\begin{align}\label{Solve_Ek1}
E_{k,1}^{*}[n]=p_{k,1}^{*}[n]t_{k,1}^{*}[n],
\end{align}
\begin{align}\label{Solve_pk1}
p_{k,1}^{*}[n]= \left[\frac{\hat{\lambda}_{k,n}B_0}{\ln2}-\frac{1}{\bar{\gamma}_0}\right]_0^{P_u^{max}} ,
\end{align}
\begin{equation}\label{Solve_tk1}
t_{k,1}^{*}[n]\left\{\begin{array}{ll}{=\delta_t,} & {\text { if } p_{k,1}^{*}[n]-\hat{\lambda}_{k,n}r_{uh}\left(p_{k,1}^{*}[n]\right)+\eta_{k,n}<0} \\
{\epsilon[0, \delta_t],} & {\text { if } p_{k,1}^{*}[n]-\hat{\lambda}_{k,n}r_{uh}\left(p_{k,1}^{*}[n]\right)+\eta_{k,n}^n=0} \\
{=0,} & {\text { if } p_{k,1}^{*}[n]-\hat{\lambda}_{k,n}r_{uh}\left(p_{k,1}^{*}[n]\right)+\eta_{k,n}>0}\end{array}\right.
\end{equation}

\end{subequations}
\end{small}
\end{lemma}
\begin{IEEEproof}
See Appendix A.
\end{IEEEproof}

The solutions to the subproblems (L2)-(L6) are given in Lemmas \ref{lemma4}-\ref{lemma8}, respectively, and the proofs of these problems are omitted here due to de similar KKT method applied in the Lemma \ref{lemma3}.

\begin{lemma}\label{lemma4}
By solving subproblem (L2) with KKT, the optimal solution can be denoted as

\begin{small}
\begin{subequations}
\begin{align}
E_{k,2}^{*}[n]=p_{k,2}^{*}[n]t_{k,2}^{*}[n],
\end{align}
\begin{align}
p_{k,2}^{*}[n]= \left[\frac{\mu_{k,n}B_0}{\ln2}-\frac{1}{\bar{\gamma}_0}\right]_0^{P_u^{max}} ,
\end{align}

\begin{equation}
t_{k,2}^{*}[n]\left\{\begin{array}{ll}{=\delta_t,} & {\text { if } p_{k,2}^{*}[n]-\mu_{k,n}r_{uh}\left(p_{k,2}^{*}[n]\right)+\eta_{k,n}<0} \\
{\epsilon[0, \delta_t],} & {\text { if } p_{k,2}^{*}[n]-\mu_{k,n}r_{uh}\left(p_{k,2}^{*}[n]\right)+\eta_{k,n}^n=0} \\
{=0,} & {\text { if } p_{k,2}^{*}[n]-\mu_{k,n}r_{uh}\left(p_{k,2}^{*}[n]\right)+\eta_{k,n}>0}\end{array}\right.
\end{equation}
\end{subequations}
\end{small}
\end{lemma}

\begin{lemma}\label{lemma5}
By solving subproblem (L3) with KKT, the optimal solution can be denoted as

\begin{small}
\begin{subequations}
\begin{align}
E_{k,3}^{*}[n]=p_{k,3}^{*}[n]t_{k,3}^{*}[n],
\end{align}
\begin{align}
p_{k,3}^{*}[n]= \left[\frac{\nu_{k,n}B_0}{\ln2}-\frac{1}{\bar{\gamma}_1}\right]_0^{P_h^{max}}, \label{remarks}
\end{align}
\begin{equation}
t_{k,3}^{*}[n]\left\{\begin{array}{ll}{=\delta_t,} & {\text { if } p_{k,3}^{*}[n]-\nu_{k,n}r_{ha}\left(p_{k,3}^{*}[n]\right)+\eta_{k,n}<0} \\
{\epsilon[0, \delta_t],} & {\text { if } p_{k,3}^{*}[n]-\nu_{k,n}r_{ha}\left(p_{k,3}^{*}[n]\right)+\eta_{k,n}=0} \\
{=0,} & {\text { if } p_{k,3}^{*}[n]-\nu_{k,n}r_{ha}\left(p_{k,3}^{*}[n]\right)+\eta_{k,n}>0}\end{array}\right.
\end{equation}
where $\bar{\gamma}_1=\frac{\gamma_1}{||\mathbf{q}[n]-\mathbf{w}_a[n]||^2+H^2}$.
\end{subequations}
\end{small}
\end{lemma}

\begin{lemma}\label{lemma6}
By solving subproblem (L4) with KKT, the optimal solution can be denoted as
\begin{small}
\begin{align}
l_{u,k}^{*}[n]= \delta_t\left[\sqrt{\frac{\omega_{k,n}}{3\kappa_uc_u^3}}\right]_0^{\frac{f_u^{max}}{c_u}} ,
\end{align}
\end{small}
\end{lemma}

\begin{lemma}\label{lemma7}
By solving subproblem (L5) with KKT, the optimal solution can be denoted as
\begin{small}
\begin{equation}
l_{h,k}^{*}[n]\left\{\begin{array}{ll}{=\delta_t\left[\sqrt{\frac{\omega_{k,n}-\hat{\lambda}_{k,n}}{3\kappa_hc_h^3}}\right]_0^{\frac{\bar{f}_{h}^{max}}{c_h}},} & {\text { if } \omega_{k,n}-\hat{\lambda}_{k,n}\geq0} \\
{=0,} & {\text { if } \omega_{k,n}-\hat{\lambda}_{k,n}<0}\end{array}\right.
\end{equation}
\end{small}
\end{lemma}

\begin{lemma}\label{lemma8}
By solving subproblem (L6) with KKT, the optimal solution can be denoted as
\begin{align}
l_{a,k}^{*}[n]\left\{\begin{array}{ll}{=0,} & {\text { if } \mu_{k,n}+\nu_{k,n}-\omega_{k,n}>0} \\
{=a,} & {\text { if } \mu_{k,n}+\nu_{k,n}-\omega_{k,n}=0}\end{array}\right.
\end{align}
where $a$ represent any non-negative constant.
\end{lemma}

Based on the duality method, it can be seen from Lemma \ref{lemma3}-\ref{lemma5} that the offloading strategy depends on the channel quality between the UAV and TDs or that between the UAV and AP. For example, the expression \eqref{remarks} indicates that the UAV would help TDs forward the task bits to AP if the distance between the UAV and AP is smaller than a threshold, i.e., $d_{ha}[n]\leq \sqrt{\frac{\nu_{k,n}B_0}{\ln2}\bar{\gamma}_1}$. Moreover,  from Lemma  \ref{lemma6} and \ref{lemma7}, we can know that TDs would choose to perform bits offloading to UAV for computing when the local computation task exceed the amount of $\sqrt{\frac{\hat{\lambda}_{k,n}}{3\kappa_hc_h^3}}\delta_t$. Otherwise, the TDs only operate local computing.

\vspace{-0.5em}
\subsection{Obtaining $\left(\boldsymbol{\lambda}, \boldsymbol{\mu},\boldsymbol{\nu},\boldsymbol{\omega},\boldsymbol{\eta} \right)$ by Solving Problem (D2)}
After obtaining $\left( \mathbf{L}^{*},\mathbf{t}^{*},\mathbf{E}^{*} \right)$ for given $\left(\boldsymbol{\lambda}, \boldsymbol{\mu},\boldsymbol{\nu},\boldsymbol{\omega},\boldsymbol{\eta} \right)$, we then can obtain the optimal dual variables by solving problem (D2), denoted by $\left(\boldsymbol{\lambda}^{*}, \boldsymbol{\mu}^{*},\boldsymbol{\nu}^{*},\boldsymbol{\omega}^{*},\boldsymbol{\eta}^{*} \right)$. Considering problem (D2) is non-differentiable in general, this motivates us to use the ellipsoid method \cite{Boyd2} to solve problem (D2). Specifically, the subgradient of the objective function can be represented by $\left(\Delta\boldsymbol{\lambda}^{T}, \Delta\boldsymbol{\mu}^{T},\Delta\boldsymbol{\nu}^{T},\Delta\boldsymbol{\omega}^{T},\Delta\boldsymbol{\eta}^{T} \right)^{T}$, in which the vectors $\Delta\boldsymbol{\lambda}, \Delta\boldsymbol{\mu},\Delta\boldsymbol{\nu},\Delta\boldsymbol{\omega},\Delta\boldsymbol{\eta}$ are respective given as

\begin{small}
\begin{subequations}
\begin{align}
\Delta\boldsymbol{\lambda}=\sum\limits_{i=1}^{n}l_{h,k}[i]-\sum\limits_{i=1}^{n}t_{k,1}[i]r_{uh}\left(\frac{E_{k,1}[i]}{t_{k,1}[i]}\right), \forall k, n,
\end{align}
\begin{align}
\Delta\boldsymbol{\mu}=l_{a,k}[n]-t_{k,2}[n]r_{uh}\left(\frac{E_{k,2}[n]}{t_{k,2}[n]}\right), \forall k, n,
\end{align}
\begin{align}
\Delta\boldsymbol{\nu}=l_{a,k}[n]-t_{k,3}[n]r_{uh}\left(\frac{E_{k,3}[n]}{t_{k,3}[n]}\right), \forall k, n,
\end{align}
\begin{align}
\Delta\boldsymbol{\omega}=L_{k,n}^{min}-l_{u,k}[n]-l_{h,k}[n]-l_{a,k}[n], \forall k, n,
\end{align}
\begin{align}
\Delta\boldsymbol{\eta}=t_{k,1}[n]+t_{k,2}[n]+t_{k,3}[n]-\delta_t, \forall k, n.
\end{align}
\end{subequations}
\end{small}
\vspace{-0.5em}
\subsection{Constructing Optimal Solution to Problem (P2)}
Due to the nonuniqueness of $t_{k,m}^{*}[n], m=\{1,2,3\}$ and $l_{a,k}^{*}[n]$, an extra step is needed to construct the optimal solution to problem (P2). From Lemma \ref{lemma3}-\ref{lemma8}, the obtained solutions $p_{k,m}^{*}[n], m=\{1,2,3\}$, $l_{u,k}^{*}[n]$, $l_{h,k}^{*}[n]$ are unique. By substituting these parameters in problem (P2), we have

\begin{small}
\begin{subequations}\label{re_construct}
\begin{align}
 \min \limits_{l_{a,k}[n],\boldsymbol{t},\mathbf{E}} &\sum\limits_{n=1}^{N}\sum\limits_{k=1}^{K} E_{k,1}[n]+E_{k,2}[n]+E_{k,3}[n]  \\
\text{s.t.} ~~&\eqref{C7}\text{-}\eqref{C12},\\
& \sum\limits_{i=1}^{n}l_{h,k}^{*}[i]\leq\sum\limits_{i=1}^{n}t_{k,1}[i]r_{uh}\left( p_{k,1}^{*}[i] \right),\forall k, n, \label{rC1}\\
&l_{a,k}[n] \leq t_{k,2}[n]r_{uh}\left( p_{k,2}^{*}[n] \right), \forall k, n,\label{rC2}\\
 &l_{a,k}[n]\leq t_{k,3}[n]r_{ha}\left( p_{k,3}^{*}[n] \right), \forall k, n, \label{rC3}\\
 &l_{u,k}^{*}[n]+l_{h,k}^{*}[n]+l_{a,k}[n]\geq L_{k,n}^{min}, \forall k, n,\label{rC4}
\end{align}
\end{subequations}
\end{small}

By solving the linear programming problem \eqref{re_construct}, the optimal solution to primal problem (P2) is obtained. The details for solving problem (P2) is summarized in Algorithm~\ref{A1}.

\begin{algorithm}

  \caption{A dual algorithm to optimally solve (P2)}
  \small
  \label{A1}
  \begin{algorithmic}[1]
    \State Initialization: $\boldsymbol{\lambda}, \boldsymbol{\mu},\boldsymbol{\nu},\boldsymbol{\omega},\boldsymbol{\eta}$, and the ellipsoid.
    \Repeat
     \State Based on Lemma \ref{lemma3}-\ref{lemma8}, obtain $\mathbf{L}^{*},\mathbf{t}^{*},\mathbf{E}^{*}$.
     \State By solving problem (D2), obtain the subgradients of the objective functions and constraints.
     \State Update $\boldsymbol{\lambda}, \boldsymbol{\mu},\boldsymbol{\nu},\boldsymbol{\omega},\boldsymbol{\eta}$ based on ellipsoid method.
   \Until  $\boldsymbol{\lambda}, \boldsymbol{\mu},\boldsymbol{\nu},\boldsymbol{\omega}$ and $\boldsymbol{\eta}$ converge.
     \State Let $\left(\boldsymbol{\lambda}^{*}, \boldsymbol{\mu}^{*},\boldsymbol{\nu}^{*},\boldsymbol{\omega}^{*},\boldsymbol{\eta}^{*}\right) \leftarrow \left(\boldsymbol{\lambda}, \boldsymbol{\mu},\boldsymbol{\nu},\boldsymbol{\omega},\boldsymbol{\eta}\right)$.
     \State Obtain $p_{k,m}^{*}[n], m=\{1,2,3\}$, $l_{u,k}^{*}[n]$, $l_{h,k}^{*}[n]$  based on  Lemma \ref{lemma3}-\ref{lemma8}, and then obtain optimal $t_{k,m}^{*}[n], m=\{1,2,3\}$ and $l_{a,k}^{*}[n]$ by solving problem \eqref{re_construct}.
  \end{algorithmic}
\end{algorithm}

\section{Energy Minimization with Trajectory Optimization}
In this section, the UAV trajectory is designed to further decrease the total energy consumption.  Based on $\left\{\mathbf{p}^{*}, \mathbf{E}^{*}, \mathbf{t}^{*}\right\}$ obtained by solving Algorithm \ref{A1}, where $\mathbf{p}^{*}$ satisfies $\mathbf{p}^{*}=\frac{\mathbf{E}^{*}}{\mathbf{t}^{*}}$, the energy minimization problem by optimizing the UAV trajectory is formulated as

\begin{subequations}
\begin{align}
& \text{(P3):}\min \limits_{\mathbf{L},\boldsymbol{Q}} \sum\limits_{n=1}^{N}\sum\limits_{k=1}^{K}  \frac{k_u\left(c_ul_{u,k}[n]\right)^3}{\delta_t^2} + \frac{k_h\left(c_hl_{h,k}[n]\right)^3}{\delta_t^2} \nonumber\\
&~~~~~~~~~~~~~~~~~~~~~~~~~~~~~~~~~~~~~~~~~~ + w\delta_t\sum\limits_{n=1}^{N}P\left(||\mathbf{v}[n]||\right)  \nonumber\\
&\text{s.t.} ~~\eqref{T1}\text{-}\eqref{T3},~\eqref{obtain_v},~ \eqref{C5}\text{-}\eqref{C7},~ \eqref{C10}, \nonumber\\
&\sum\limits_{i=1}^{n}l_{h,k}[i]\leq\sum\limits_{i=1}^{n}t_{k,1}^{*}[i]B_0\log_2\left( 1+\frac{p_{k,1}^{*}[i]\gamma_0}{||\mathbf{q}[i]-\mathbf{w}_k||^2+H^2} \right),\nonumber\\
&~~~~~~~~~~~~~~~~~~~~~~~~~~~~~~~~~~~~~~~~~~~~~~ \label{C_UAV1}\\
&l_{a,k}[n] \leq t_{k,2}^{*}[n]B_0\log_2\left( 1+\frac{p_{k,2}^{*}[n]\gamma_0}{||\mathbf{q}[n]-\mathbf{w}_k||^2+H^2} \right),\label{C_UAV2}\\
 &l_{a,k}[n]\leq t_{k,3}^{*}[n]B_0\log_2\left( 1+\frac{p_{k,3}^{*}[n]\gamma_0}{||\mathbf{q}[n]-\mathbf{w}_a||^2+H^2} \right). \label{C_UAV3}
\end{align}
\end{subequations}

It can be seen from problem (P3) that the objective function is non-convex and  the expressions in \eqref{C_UAV1}-\eqref{C_UAV3} are non-convex with respect to $\mathbf{q}[n]$. Hence, problem (P3) belongs to a non-convex optimization problem that is challenging to be solved. To tackle the non-convexity, the SCA technique is applied.

 To tackle the non-convexity of the function $P\left(||\mathbf{v}[n]||\right)$ in the objective function, we first introduce the slack variable $v_n\geq||\mathbf{v}[n]||$, thus the expression in \eqref{p_consumption} can be rewritten as
\begin{align}
&P(v_n)=P_0\left(1+\frac{3v^2_n}{U^2_{tip}}\right)+\nonumber\\
&~~~~P_i\left(\sqrt{1+\frac{v_n^4}{4v_0^4}}-\frac{v_n^2}{2v_0^2}\right)^{\frac{1}{2}}+\frac{1}{2}d_0\rho sAV_n^3, \forall n\in\mathcal{N}.\label{cvx_E}
\end{align}
~~Note that the second term in \eqref{cvx_E} is still non-convex. By introducing another slack variable $u_n^2\geq \sqrt{1+\frac{v_n^4}{4v_0^4}}-\frac{v_n^2}{2v_0^2}$, we can readily obtain the expression as
\begin{align}
\frac{1}{u_n^2}\leq u_n^2+\frac{v_n^2}{v_0^2},\forall n\in\mathcal{N},\label{CVX_E2}
\end{align}
then for given any local point $\{v_{n,j},u_{n,j}\}$ ($j$ denotes $j$th iteration), the right-hand side (RHS) of \eqref{CVX_E2} can be lower-bounded  via the first-order Taylor expansion as it is jointly convex with respect to $v_n$ and $u_n$ \cite{Boyd}. Let $\chi_n^{lb}$ denote this  lower bound function which is expressed as
\begin{align}
\chi_n^{lb}&\triangleq \left(u_{n,j}\right)^2+2u_{n,j}\left(u_n-u_{n,j}\right)+\left(v_{n,j}\right)^2\frac{1}{v_0^2}+ \nonumber\\
       &\quad\frac{2v_{n,j}}{v_0^2}\left(v_n-v_{n,j}\right), \forall n\in\mathcal{N}.\label{cvx_E3}
\end{align}
~~Based on the discussion above, the UAV's flight energy consumption can be approximately expressed as a convex function, i.e.,
\begin{align}
P_{appro}(v_n)=P_0\left(1+\frac{3v^2_n}{U^2_{tip}}\right)+P_iu_n+\frac{1}{2}d_0\rho sAv_n^3
\end{align}

Considering the expression of $\log_2\left( 1+\frac{p_{k,1}^{*}[i]\gamma_0}{||\mathbf{q}[i]-\mathbf{w}_k||^2+H^2} \right)$ in RHS of \eqref{C_UAV1}, it is non-convex with respect to $\mathbf{q}[i]$. However, it can be still deemed as a convex expression if taking $||\mathbf{q}[i]-\mathbf{w}_k||^2$ as a whole.  Hence, for any given local point $\{\mathbf{q}_j[n]\}$, the lower bound function of the RHS of expression in \eqref{C_UAV1} can be denoted by $\varphi_{k,1}^{lb}[i]$, given in \eqref{la_computation2}.

Similarly, for any given local point $\{\mathbf{q}_j[n]\}$, the  RHSs of the inequalities in  \eqref{C_UAV2} and \eqref{C_UAV3} can be also lower-bounded. The corresponding lower bound functions can be derived, as expressed in \eqref{la_computation2} and \eqref{la_computation3}, respectively.

\newcounter{mytempeqncnt2}
\begin{figure*}[!t]
\begin{small}
\begin{align}\label{la_computation2}
\varphi_{k,1}^{lb}[i]= \log_2\left( 1+\frac{p_{k,1}^{*}[i]\gamma_0}{||\mathbf{q}_j[i]-\mathbf{w}_k||^2+H^2}\right)-\frac{\log_2(e)p_{k,1}^{*}[i]\gamma_0\left(||\mathbf{q}[i]-\mathbf{w}_k||^2-||\mathbf{q}_j[i]-\mathbf{w}_k||^2\right)}
{(||\mathbf{q}_j[i]-\mathbf{w}_k||^2+H^2)(||\mathbf{q}_j[i]-\mathbf{w}_k||^2+H^2+p_{k,1}^{*}[i]\gamma_0)}.
\end{align}
\end{small}
\end{figure*}

\newcounter{mytempeqncnt3}
\begin{figure*}[!t]
	\begin{small}
\begin{align}\label{la_computation3}
\varphi_{k,2}^{lb}[n]= \log_2\left( 1+\frac{p_{k,2}^{*}[n]\gamma_0}{||\mathbf{q}_j[n]-\mathbf{w}_k||^2+H^2}\right)-\frac{\log_2(e)p_{k,2}^{*}[i]\gamma_0\left(||\mathbf{q}[n]-\mathbf{w}_k||^2-||\mathbf{q}_j[n]-\mathbf{w}_k||^2\right)}
{(||\mathbf{q}_j[n]-\mathbf{w}_k||^2+H^2)(||\mathbf{q}_j[n]-\mathbf{w}_k||^2+H^2+p_{k,2}^{*}[n]\gamma_0)}.
\end{align}
\end{small}
\end{figure*}

\newcounter{mytempeqncnt4}
\begin{figure*}[!t]
	\begin{small}
\begin{align}\label{la_computation4}
\varphi_{k,3}^{lb}[n]= \log_2\left( 1+\frac{p_{k,3}^{*}[n]\gamma_1}{||\mathbf{q}_j[n]-\mathbf{w}_a||^2+H^2}\right)-\frac{\log_2(e)p_{k,3}^{*}[i]\gamma_1\left(||\mathbf{q}[n]-\mathbf{w}_a||^2-||\mathbf{q}_j[n]-\mathbf{w}_a||^2\right)}
{(||\mathbf{q}_j[n]-\mathbf{w}_a||^2+H^2)(||\mathbf{q}_j[n]-\mathbf{w}_a||^2+H^2+p_{k,3}^{*}[n]\gamma_0)}.
\end{align}
\end{small}
\hrulefill
\end{figure*}

By replacing the derived lower bound functions  and the approximately convex expression  into problem (P3), we can obtain
\begin{subequations}
 \begin{align}
 \text{(P3.1):}\min \limits_{\mathbf{L},\boldsymbol{Q} ,v_n,u_n} \sum\limits_{n=1}^{N}\sum\limits_{k=1}^{K} & \frac{k_u\left(c_ul_{u,k}[n]\right)^3}{\delta_t^2} + \frac{k_h\left(c_hl_{h,k}[n]\right)^3}{\delta_t^2}    \nonumber\\
&~~~~~ + w\delta_t\sum\limits_{n=1}^{N}P_{appro}(v_n)  \nonumber
 \end{align}
 \begin{align}
&\text{s.t.} ~~\eqref{T1}\text{-}\eqref{T3},~\eqref{obtain_v},~\eqref{C5}\text{-}\eqref{C7},~ \eqref{C10}, \nonumber\\
&\sum\limits_{i=1}^{n}l_{h,k}[i]\leq\sum\limits_{i=1}^{n}t_{k,1}^{*}[i]B_0\varphi_{k,1}^{lb}[i], \forall k, n, \label{CC_UAV1}\\
&l_{a,k}[n] \leq t_{k,2}^{*}[n]B_0\varphi_{k,2}^{lb}[n], \forall k, n,\label{CC_UAV2}\\
 &l_{a,k}[n]\leq t_{k,3}^{*}[n]B_0\varphi_{k,3}^{lb}[n], \forall k, n. \label{CC_UAV3}\\
 & v_n^2\geq||\mathbf{v}[n]||^2, n\in\mathcal{N},\label{cvx_E4}\\
&\chi_n^{lb}\geq\frac{1}{u_n^2}, n\in\mathcal{N}.\label{cvx_E5}
\end{align}
\end{subequations}

It can be readily proved that the optimal solution always makes equality hold in \eqref{CC_UAV2}, \eqref{CC_UAV3} and \eqref{cvx_E5}. Also, the equality must holds in the causality condition \eqref{CC_UAV1} for $n=N$. Hence, the problem (P3.1) is equivalent to (P3). Obviously, Problem (P3.1) is convex that can be solved by standard convex optimization tools, such as CVX \cite{Boyd3}.

In summary, a overall iterative algorithm that jointly optimizes computation bits allocation, power allocation, time slot scheduling and UAV trajectory can be derived to solve the primal problem (P1), as summarized in Algorithm 2. Algorithm 2 consists of the duality method and SCA technology, at least a locally optimal solution always can be achieved by the proposed joint optimization algorithm.
\begin{algorithm}
  \caption{The overall iterative algorithm to solve (P1)}
  \small
  \label{A2}
  \begin{algorithmic}[1]
    \State Given UAV initial local point $\{\mathbf{q}_j[n]\}$, $\{v_{n,j}\}$ and $\{u_{n,j}\}$, let iteration $j=0$.
    \Repeat
     \State With $\{\mathbf{q}_j[n]\}$, solve (P2) based on Algorithm 1 and obtain $\left\{\mathbf{p}^{*}, \mathbf{E}^{*}, \mathbf{t}^{*}\right\}$.
     \State With $\left\{\mathbf{p}^{*}, \mathbf{E}^{*}, \mathbf{t}^{*}\right\}$ and $\{\mathbf{q}_j[n]\}$, solve  problem (P3.1), and obtain optimized trajectory denoted by $\{\mathbf{q}_j^{*}[n]\}$, $\{v_{n,j}^{*}\}$ and $\{u_{n,j}^{*}\}$ via CVX.
     \State Update $\{\mathbf{q}_{j+1}[n]\}\leftarrow\{\mathbf{q}_j^{*}[n]\}$, $\{v_{n,j}\}\leftarrow\{v_{n,j}^{*}\}$,  $\{u_{n,j}\}\leftarrow\{u_{n,j}^{*}\}$.
     \State Update $j\leftarrow j+1$.
   \Until  The objective value converges.
  \end{algorithmic}
\end{algorithm}

Here, we briefly give the complexity analysis for the proposed algorithms. For each iteration of Algorithm \ref{A2}, it consists of solving Algorithm \ref{A1} and optimizing UAV trajectory with CVX. The computation complexity of Algorithm \ref{A1} mainly depends on the loop, i.e., step 3) to step 5) of Algorithm \ref{A1}. Note that the complexity of ellipsoid method is  $\mathcal{O}(K^2N^2)$ \cite{Boyd}\cite{Boyd2}. Thus the complexities of step 3), 4) and 5) of Algorithm \ref{A1} are $\mathcal{O}(KN)$, $\mathcal{O}(KN)$ and $\mathcal{O}(K^2N^2)$, respectively. As a result, the total complexity for the Algorithm \ref{A1} is $\mathcal{O}(K^4N^4)$.

\vspace{-0.5em}
\section{Numerical Results}
In this section, the numerical results are presented to validate our proposed design. The vector $\mathbf{L}_{m}\in\mathbb{R}^{1\times3}$ is utilized to represent the set of required computation bits, in which the $k$th entry  stands for the required computation task for TD $k$ in per time slot. The details of parameter setup are shown in Table \ref{table1}.

\begin{table}[h]
\caption{System Parameters for Numerical Simulation}
\label{table1}
\setlength{\tabcolsep}{3pt}
\begin{tabular}{|p{145pt}|p{95pt}|}
\hline
Symbolic Meaning&
Symbol and Value\\
\hline
Altitude of UAV&
$H=20$ m\\
Amount of TDs&
$K=3$\\
Maximum speed&
$V_{max}=20$ m/s\\
Initial location of UAV&
$\mathbf{q}_0=[-20,-20]$ m\\
Final location of UAV&
$\mathbf{q}_F=[20,-20]$ m\\
Time slot size&
$\delta_t=0.2$ s\\
Maximum instantaneous power of each users for offloading&
$P_{u}^{max}=35$ dBm\\
Maximum instantaneous power of UAV&
$P_{h}^{max}=35$ dBm\\
Noise power spectrum density&
$N_0=N_1=-130$ dBm/Hz\\
Reference channel power&
$\beta_0=-50$ dB\\
Communication bandwidth&
$B=10$ MHz\\
Maximum CPU frequency of each TD&
$f_u^{max}=2$ GHz\\
Maximum CPU frequency of UAV&
$f_h^{max}=3$ GHz\\
Required CPU cycles per bit computation at TD&
$c_u=10^3$ cycles/bit\\
Required CPU cycles per bit computation at UAV&
$c_h=10^3$ cycles/bit\\
CPU capacitance coefficient of each TD&
$k_u=10^{-27}$\\
CPU capacitance coefficient of UAV&
$k_h=10^{-27}$\\
Weight&
$w=0.01$\\
Tip speed of the rotor blade&
$U_{tip}=120$ $m/s$\\
Rotor disc area&
$A=0.503$ $m^2$\\
Air density&
$\rho=1.225$ $(kg/m^3)$\\
Rotor solidity&
$s=0.05$\\
Fuselage drag ratio&
$d_0=0.3$\\
Mean rotor induced velocity in hover&
$v_0=4.03$\\
Blade profile power in hovering status&
$P_0=158.76$ w\\
Induced power in hovering status&
$P_i=88.63$ w\\
\hline
\end{tabular}
\label{tab1}
\end{table}

\begin{figure}[htbp]
\vspace{-0.2cm}
\setlength{\abovecaptionskip}{0.cm}
\centering
\includegraphics[width=0.4\textwidth]{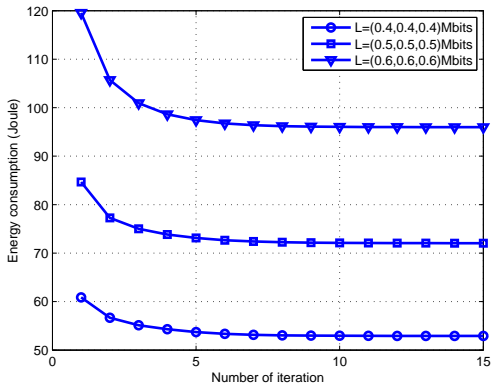}
\caption{The convergence of the proposed algorithm for period $T=6$ s.}\label{convergence_T6}
\setlength{\belowcaptionskip}{-0.cm}
\vspace{-0.5em}
\end{figure}

\begin{figure*}[htbp]
	\vspace{-0.2cm}
	\setlength{\abovecaptionskip}{0.cm}
	\begin{minipage}[t]{0.35\linewidth}
		\centering
		\includegraphics[height=4.2cm,width=5.2cm]{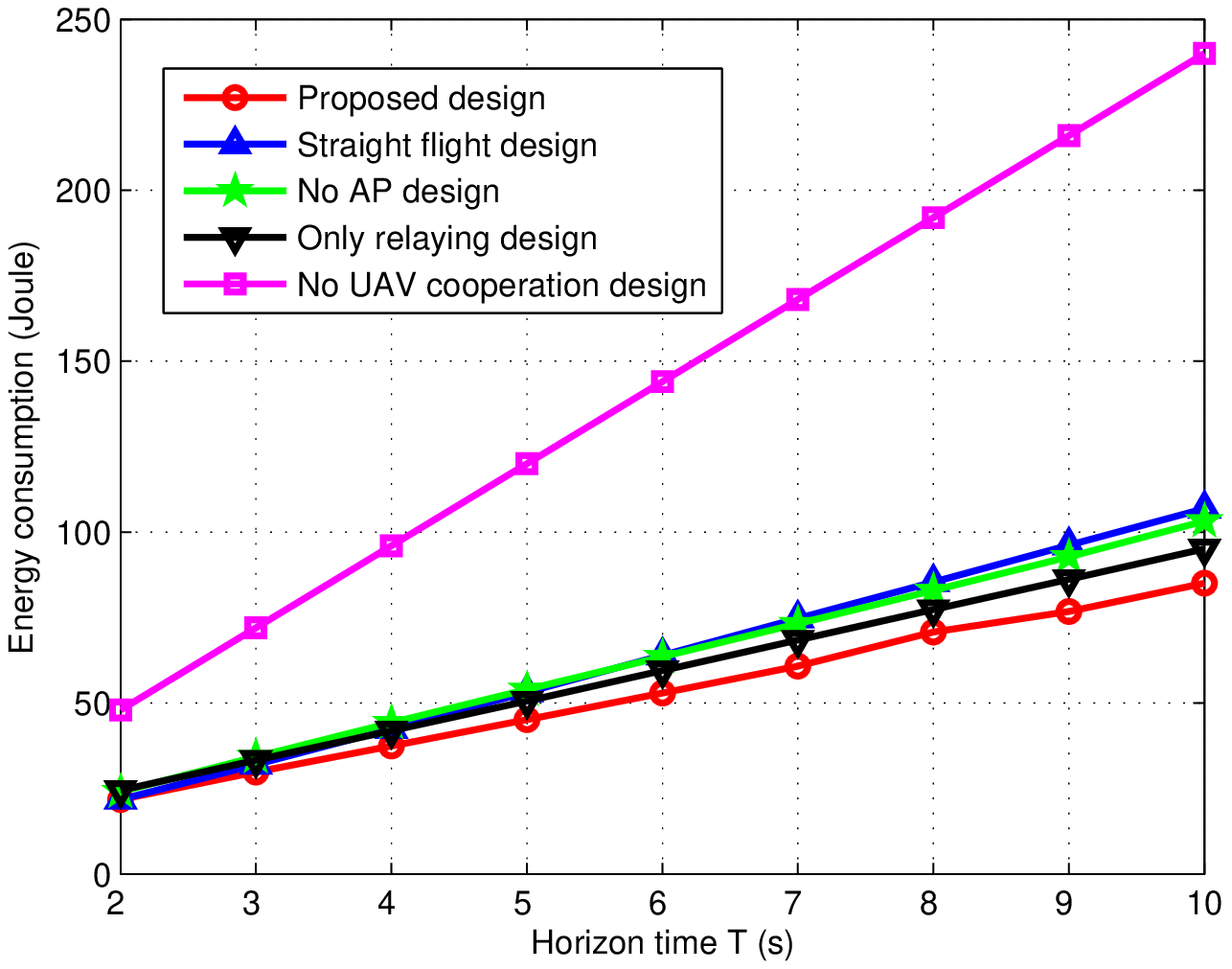}
		\caption{Energy consumption vs. period $T$.}\label{E_vs_T}
	\end{minipage}%
	\begin{minipage}[t]{0.35\linewidth}
		\centering
		\includegraphics[height=4.2cm,width=5.2cm]{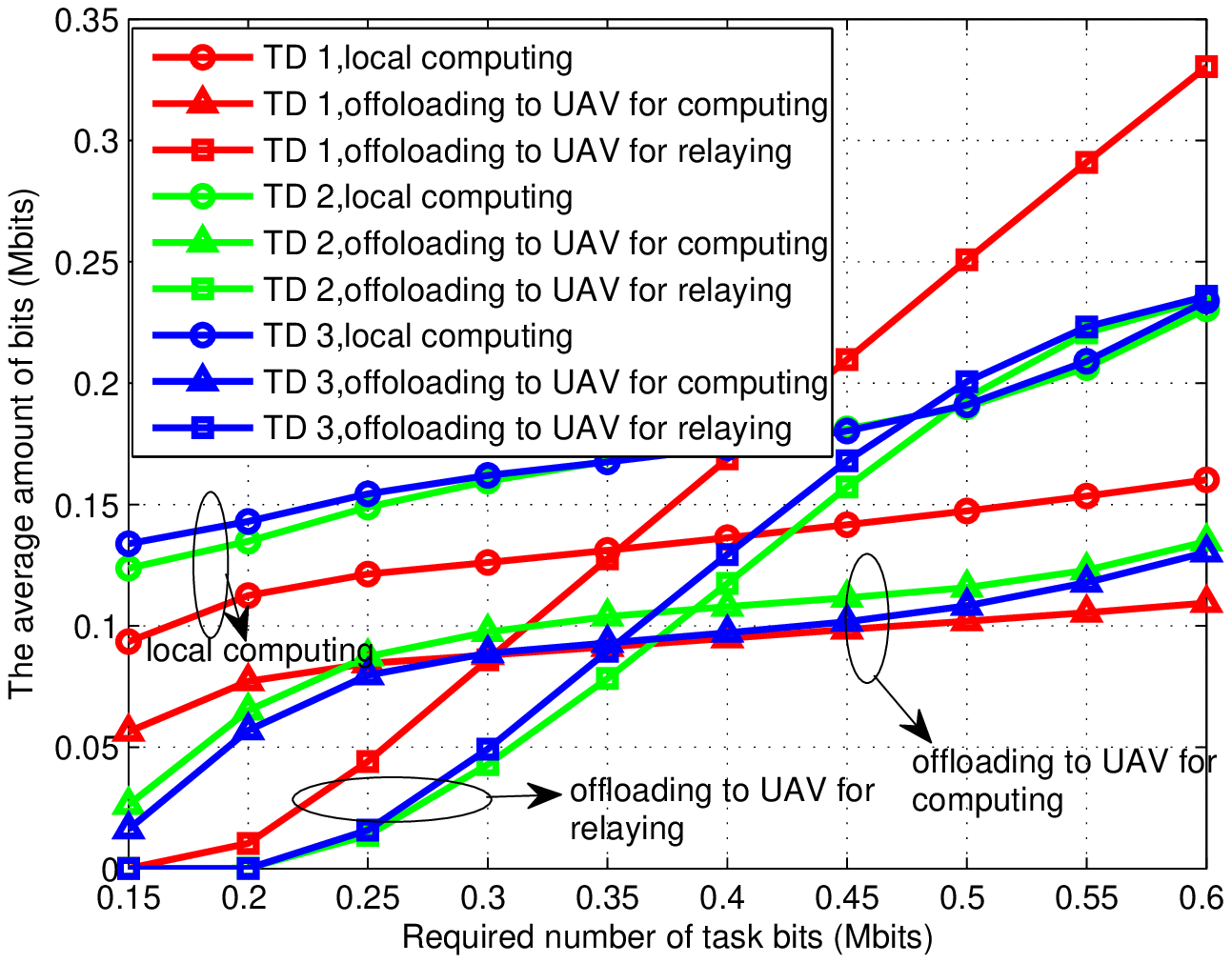}
		\caption{Average computing bits vs. task requirement.}\label{lu_lh_la}
	\end{minipage}\begin{minipage}[t]{0.35\linewidth}
		\centering
		\includegraphics[height=4.2cm,width=5.2cm]{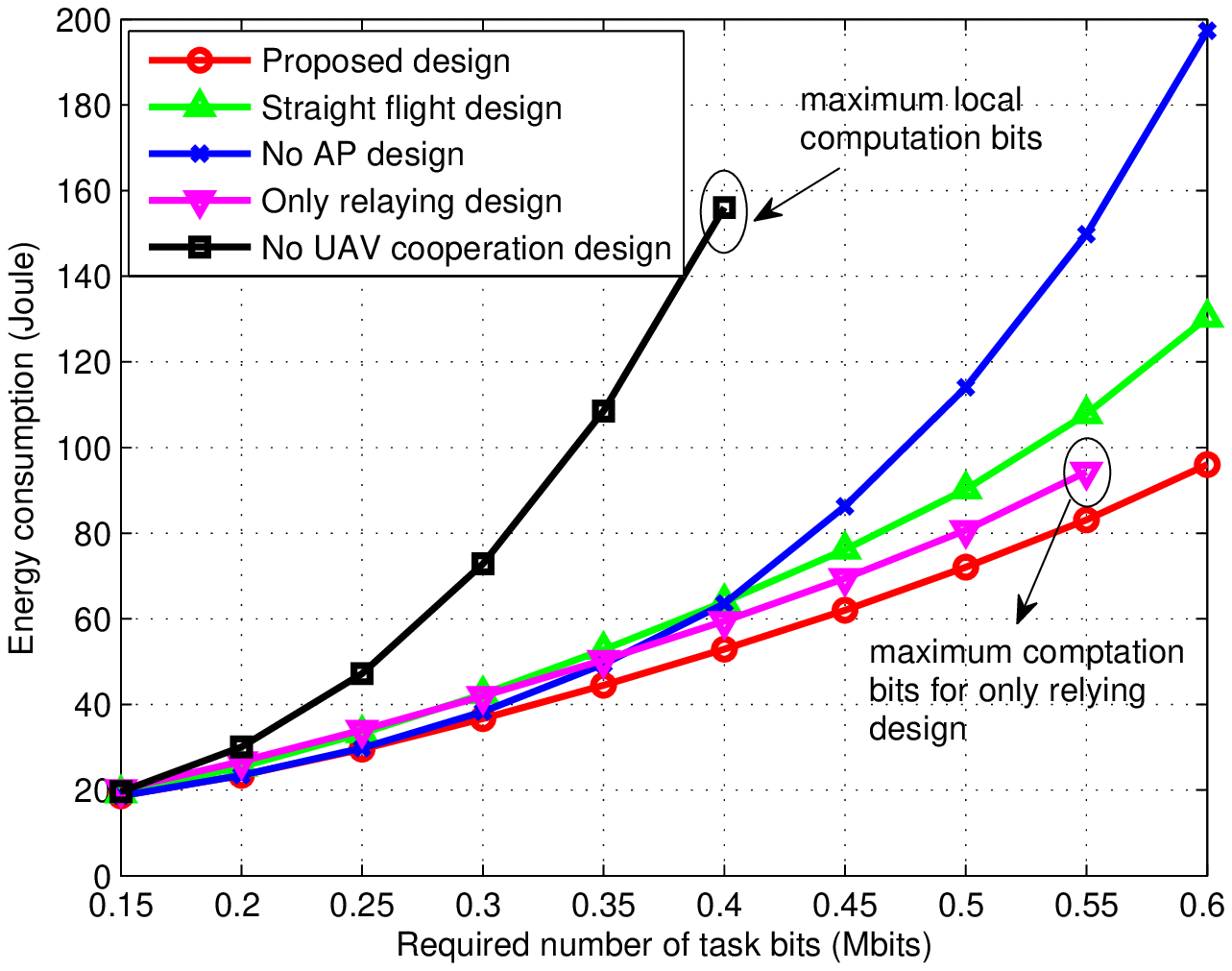}
		\caption{Energy consumption vs. task requirement.}\label{E_vs_L}
	\end{minipage}
\end{figure*}

Note that in order to illustrate the effectiveness of our proposed design, several other benchmark cases are designed as follows.
\begin{enumerate}
  \item Straight Flight design. In this case, the UAV flies from the given initial location to final location following a straight trajectory.
  \item No AP design. In this case, the task bits of TDs are computed without AP coorperation.
  \item Only Relaying design. In this case, the UAV can only act as a relay to assist task bits transmit from TDs to AP.
  \item No UAV Cooperation design. In this case, the task bits can only be computed locally at each TD. Note that for the convenience of fair comparison with the other designs, the minimum UAV's flight power consumption with the maximum-endurance speed $V_{me}$, as described in \cite{E_model}, is adopted in this design.
\end{enumerate}

Fig.~\ref{convergence_T6} shows the convergence performance of the proposed Algorithm 2, in which three cases with different computation requirements are given to compare under the period $T=6$~s. This figure shows that the proposed algorithm is guaranteed to converge nearly within 15~iterations, indicating that the proposed algorithm is highly efficient.

Fig. \ref{E_vs_T} shows the total energy consumption including communication-related energy and computation-related energy as well as the weighted UAV flight energy versus the period $T$ for task requirement $\mathbf{L}_{m}=(0.4,0.4,0.4)$ Mbits. It can be observed that the energy costed by No UAV cooperation design increases sharply with $T$ increasing. The other designs achieve more smaller value of energy consumption compared with No UAV cooperation design, this is because that the UAV as a helper can help bits offloading. In addition, it can be also observed that the proposed design always outperforms the other designs due to joint computation and communication design as well as trajectory optimization.

Fig. \ref{lu_lh_la} illustrates the average bits that are respectively computed at each TD, UAV and AP for different task requirement during  $T=6$~s. Besides, for convenience of analysis, the number of the required task bits for each TD is same, and the value is shown at $x$-coordinate axis in this picture. From Fig. \ref{lu_lh_la}, it can be seen that the AP is not necessary to join to help computation for TDs at a small value of task requirement, because the computation ability of TDs and UAV is sufficient to deal with. With the value of task requirement increasing, the UAV would tend to transmit part of computation bits to AP at the cost of certain time and energy, this is deserved and reasonable especially for a large value of task requirement since it can help release much computation resources of both TDs and UAV, so as to reduce the total energy consumption.

Fig. \ref{E_vs_L} shows the total energy consumption versus different required task bits under $T=6$~s. It is observed that the proposed design always achieves the best performance compared with other designs, and the advantages of our proposed design becomes much more evident with the value of task requirement of each TD increasing. In addition, we can find that  the No UAV Cooperation design is subject to a maximum computation ability obtained by $\frac{\delta_tf_u^{max}}{c_u}$ in \eqref{lu_constraint}. It is worth noting that  the No Relaying design is also subject to a maximum computation ability (infeasible for requirement of 0.6 Mbits/TD/slot as shown in this picture). For the No AP design, it can be seen that it costs a large amount of energy for a large value of task requirement, there are two main reasons. First, in this design the AP does not help compute; and the second is because that with the task bits increasing, the energy consumption increases with regard to the cube of required task bits, as shown in \eqref{E_u} and \eqref{E_h}. Last but not the least, due to the trajectory pattern is fixed, the Straight Flight design is limited on mobility exploitation compared with the proposed design, which causes a lager energy consumption.

\begin{figure}
	\setlength{\abovecaptionskip}{0.cm}
	\setlength{\belowcaptionskip}{-0.cm}
	\centering
	\subfigure[]{\includegraphics[width=0.25\textwidth,height=1.3in]{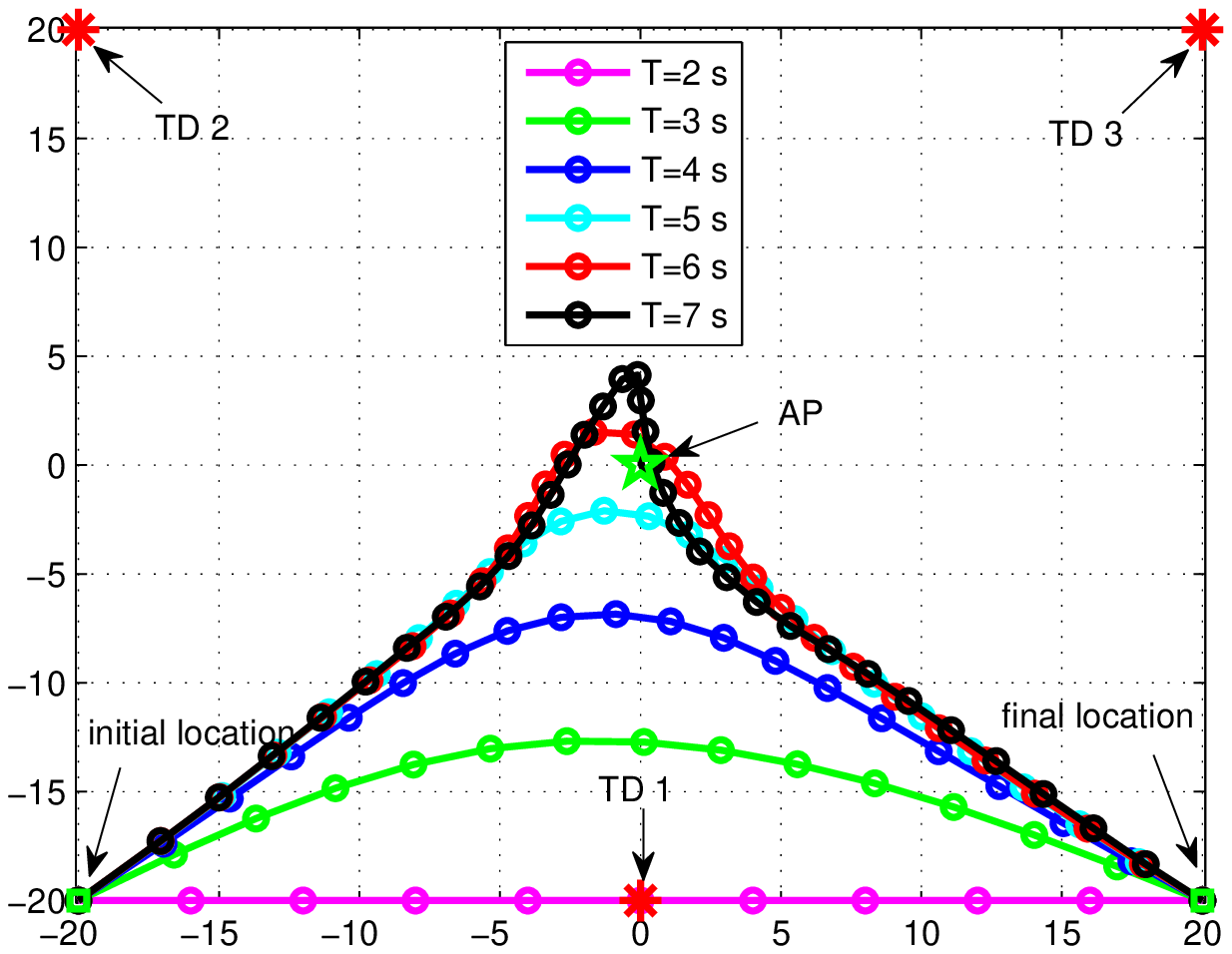}}~~~\label{tra_different_T}
	\subfigure[]{\includegraphics[width=0.25\textwidth,height=1.3in]{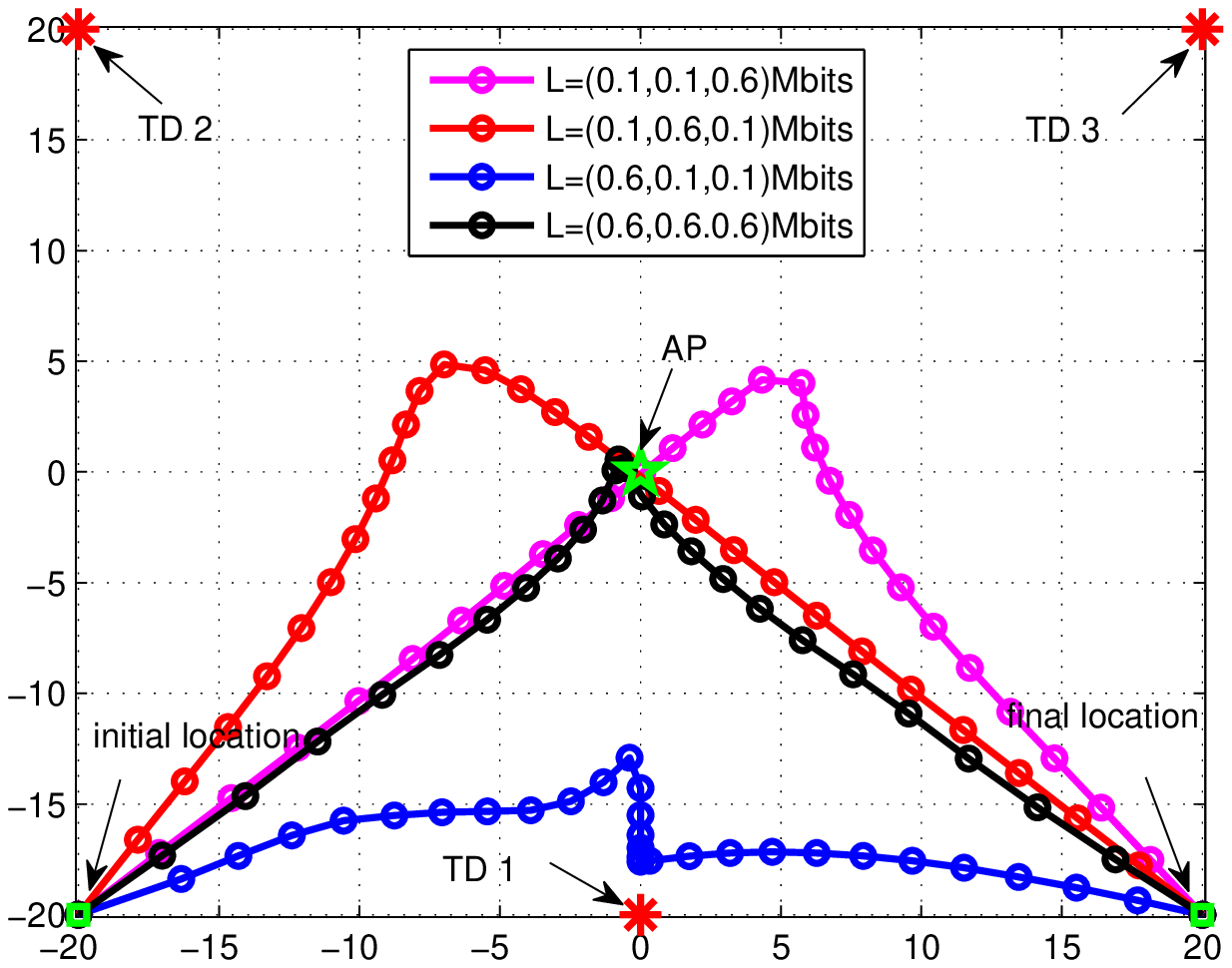}} \label{tra_different_L}
	\caption{Optimized UAV trajectory: (a). Different period T under $\mathbf{L}_{m}=(0.4,0.4,0.4)$ Mbits; (b) Different task requirement $\mathbf{L}_{m}$ under $T=6$~s } \label{trajecory}
\end{figure}

Fig. \ref{trajecory} (a) plots the UAV trajectory for different period $T$ under fixed task requirement. It can be observed that, with the value of $T$ increases, the UAV can exploit its mobility so as to seek for the optimal location in each time slot. Furthermore, it also can be observed that for a small period (e.g., $T=3$~s), the trajectory tends to be in proximity to TD 2 and TD 3 so that enhance the communication links. While for a larger period (e.g., $T=6$~s and $T=7$~s), the UAV trajectory would tend to be stable, it first flies with maximum speed and then slows down, even  tends to hover over a fixed point that can optimally balance the relationship between local computing and bits offloading.
Fig.~\ref{trajecory} (b) plots the UAV trajectory  for different task requirement under the period $T=6$~s. It can be seen that the number of required computation bits for each TD has a great effect on the UAV trajectory exploitation. Intuitively, the UAV always flies closer to the TD with high demand for computing. This is readily comprehended  that the TD with large numbers of required computation bits is eager to offload its computation bits to UAV for computing or relaying, hence the UAV should fly closer to the user so as to reduce the pathloss.

\begin{figure}
\vspace{-0.2cm}
\setlength{\abovecaptionskip}{0.cm}
\centering
\includegraphics[width=0.4\textwidth]{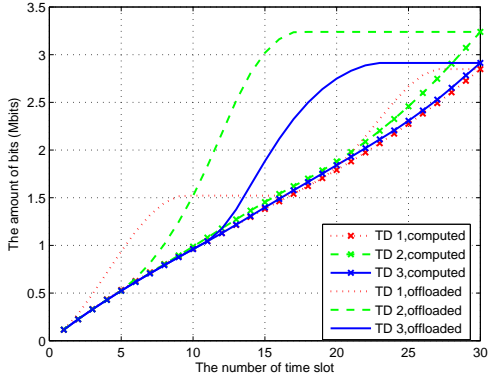}
\caption{Accumulated task bits offloaded and computed by UAV for each TD.}\label{C_Lh}
\setlength{\belowcaptionskip}{-0.cm}
\vspace{-1em}
\end{figure}

In Fig.~\ref{C_Lh}, curves about the accumulated numbers of bits computed by the UAV and that offloaded to the UAV for computing for each TD  are plotted. The required task and period in this case are set as $\mathbf{L}_{m}=(0.4,0.4,0.4)$ Mbits and $T=6$ s, respectively. From this picture, it is interesting to observe that each TD offloads its task bits to the UAV  deciding on the channel quality and task requirement. Specifically, at the beginning, the UAV is more closer to TD 1, then TD 1 offloads great numbers of task bits to it. When the UAV flies closer to TD 2 and TD 3, it receives much more bits offloaded by TD 2 and TD 3, during which the TD 1 would  reduce the offloading rate or even stop offloading for the purpose of releasing more time resource for TD 2 and TD 3, until the numbers of computed bits by the UAV are accumulated near to the sum of offloaded bits before. This mechanism can make the system resources including communication resource and computation resource utilized sufficiently. What's more, from Fig. \ref{C_Lh}, we can see that at the last time slot, i.e., $n=N$, the total computed bits for each TD at the UAV equals to the total received bits offloaded from the TD, which validates that the equality must hold in \eqref{lh_computation} for $n=N$.

\vspace{-0.5em}
\section{Conclusion}
In this paper, we investigated a new UAV-assisted MEC system, in which the UAV could  help computing the latency-critical task bits offloaded by TDs. Also, the UAV was able to act as a relay to help computation bits  offload from TDs to AP. The sum of communication-related and computation-related  energy as well as the UAV's flight energy  was minimized by jointly optimizing the computation bits allocation, time slot scheduling, power allocation and  UAV trajectory.
The proposed problem was decomposed into two subproblems that were solved by the Lagrangian duality method and SCA technique, respectively. Then, an iterative algorithm was proposed to solve the primal problem.
The numerical results validated the effectiveness of our proposed algorithm and showed the superiority of our proposed design, as compared to the other benchmark designs.
\vspace{-0.5em}

\begin{appendices}\label{appendix A}
      \section{Proof of Lemma \ref{lemma3}  }
The Lagrangian of subproblem (L1) is given as
\begin{small}
\begin{align}
&\mathcal{L}_1(\Xi)= E_{k,1}[n] - \hat{\lambda}_{k,n}t_{k,1}[n]r_{uh}\left( \frac{E_{k,1}[n]}{t_{k,1}[n]} \right)+ \eta_{k,n}t_{k,1}[n]\nonumber\\
&-aE_{k,1}[n]
+b\left(E_{k,1}[n]-t_{k,1}[n]P_{u}^{max}\right)- ct_{k,1}[n]+d\left(t_{k,1}[n]-\delta_t\right)
\end{align}
\end{small}
where $\Xi$ is the set denoted by $\Xi=(a_{k,1}^n,b_{k,1}^n,c_{k,1}^n,d_{k,1}^n)$, with $a_{k,1}^n$, $b_{k,1}^n$, $c_{k,1}^n$ and $d_{k,1}^n$ representing the non-negative Lagrange multipliers  with regard to the $E_{k,1}[n]\geq0$, $E_{k,1}[n]\leq t_{k,1}[n]P_{u}^{max}$, $t_{k,1}[n]\geq0$ and $t_{k,1}[n]\leq \delta_t$, respectively. Thus, the derivations of $\mathcal{L}_1(\Xi)$ with respect to $E_{k,1}[n]$ can be  expressed as
\begin{align}\label{L_proof_1}
\frac{\partial\mathcal{L}_1(\Xi)}{\partial E_{k,1}[n]}=1-\frac{\hat{\lambda}_{k,n}B_0}{\ln2}\frac{\bar{\gamma}_0}{1+\frac{E_{k,1}[n]\bar{\gamma}_0}{t_{k,1}[n]}}-a_{k,1}^n+b_{k,1}^n
\end{align}

Based on KKT, the complementary slackness conditions are given by $a_{k,1}^nE_{k,1}[n]=0$, $b_{k,1}^n\left(E_{k,1}[n]-t_{k,1}[n]P_{u}^{max}\right)=0$, $c_{k,1}^nt_{k,1}[n]=0$ and $d_{k,1}^n\left(t_{k,1}[n]-\delta_t\right)=0$. Let the derivation $\frac{\partial\mathcal{L}_1(\Xi)}{\partial E_{k,1}[n]}=0$, we can obtain the equation \eqref{Solve_Ek1} and \eqref{Solve_pk1}. By substituting \eqref{Solve_Ek1} into subproblem (L1) the optimal $t_{k,1}^{*}[n]$ can be easily obtained. Hence, the Lemma is proved.
 \end{appendices}


\begin{IEEEbiography}[{\includegraphics[width=1in,height=1.25in,clip,keepaspectratio]{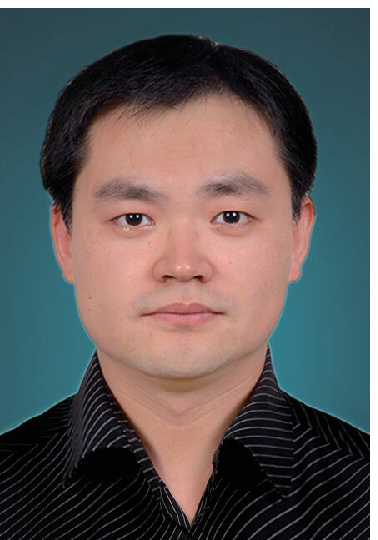}}]{Tiankui Zhang} (M'10-SM'15) received the Ph.D. degree in Information and Communication Engineering and B.S. degree in Communication Engineering from Beijing University of Posts and Telecommunications (BUPT), China, in 2008 and 2003, respectively. Currently, he is an Associate Professor in School of Information and Communication Engineering at BUPT. His research interests include wireless communication networks, mobile edge computing and caching, signal processing for wireless communications, content centric wireless networks. He had published more than 100 papers including journal papers on IEEE Journal on Selected Areas in Communications, IEEE Transaction on Communications, etc., and conference papers, such as IEEE GLOBECOM and IEEE ICC.
\end{IEEEbiography}
\vspace{-2cm}

\begin{IEEEbiography}[{\includegraphics[width=1in,height=1.25in,clip,keepaspectratio]{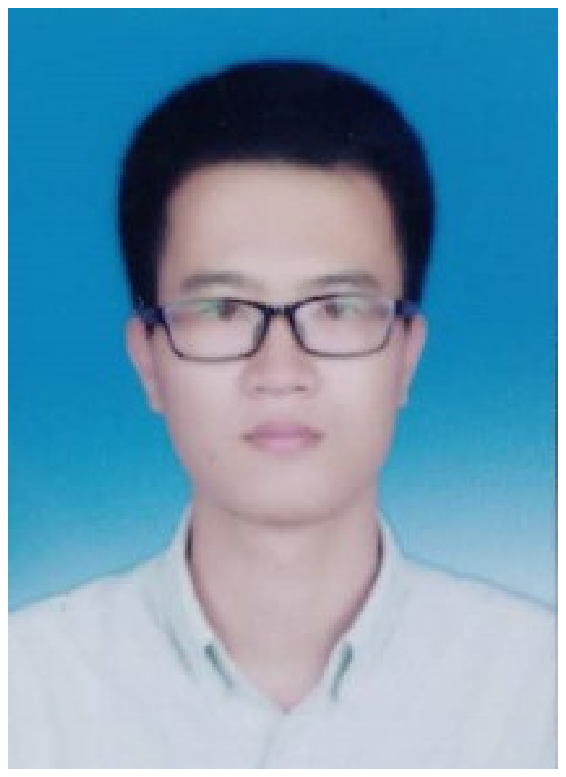}}] {Yu Xu}
received his B.S. degree from the Information Engineering School, Jiangxi University of Science and Technology, Ganzhou, China, in 2015. He received his M.S. degree from the Information Engineering School, Nanchang University, Nanchang, China, in 2019. He currently pursues his PhD degree with the School of Information and Communication Engineering, Beijing University of Posts and Telecommunications, Beijing, China. His research interests include mobile edge computing, UAV communications and wireless resource management.
\end{IEEEbiography}
\vspace{-2cm}
\begin{IEEEbiography}[{\includegraphics[width=1in,height=1.25in,clip,keepaspectratio]{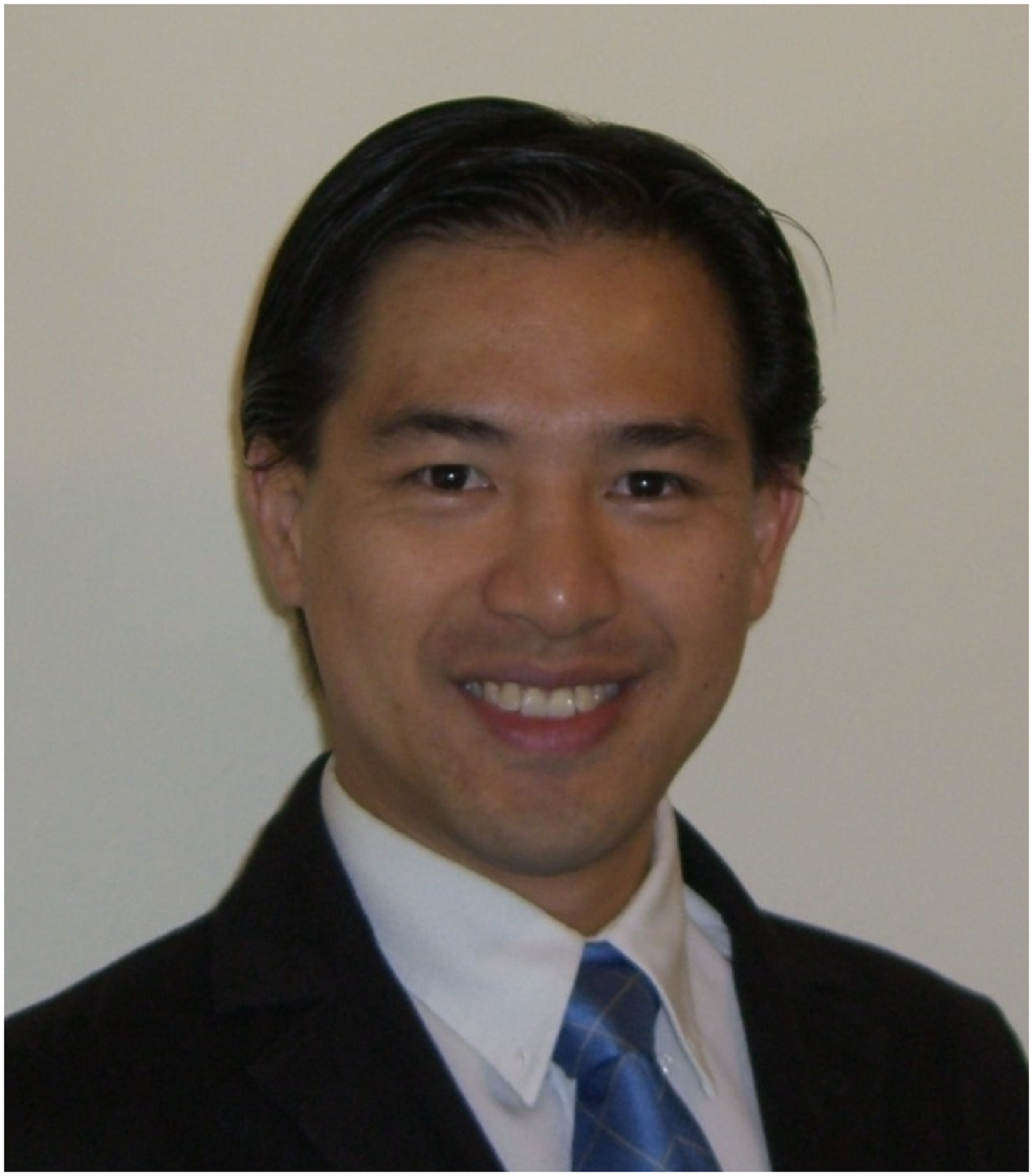}}]{Jonathan Loo} received his M.Sc. degree in Electronics (with Distinction) and the Ph.D. degree in Electronics and Communications from the University of Hertfordshire, Hertfordshire, U.K., in 1998 and 2003, respectively. Between 2003 and 2010, he was a Lecturer in Multimedia Communications with the School of Engineering and Design, Brunel University, Uxbridge, U.K.  Between June 2010 and May 2017, he was an Associate Professor in Communication Networks at the School of Science and Technology, Middlesex University, London, U.K. From June 2017, he is a Chair Professor in Computing and Communication Engineering at the School of Computing and Engineering, University of West London, United Kingdom. His research interests include machine learning and AI, information centric networking, wireless/mobile networks, network security, wireless communications, IoT/cyber-physical systems.
\end{IEEEbiography}
\vspace{-2cm}
\begin{IEEEbiography}[{\includegraphics[width=1in,height=1.25in,clip,keepaspectratio]{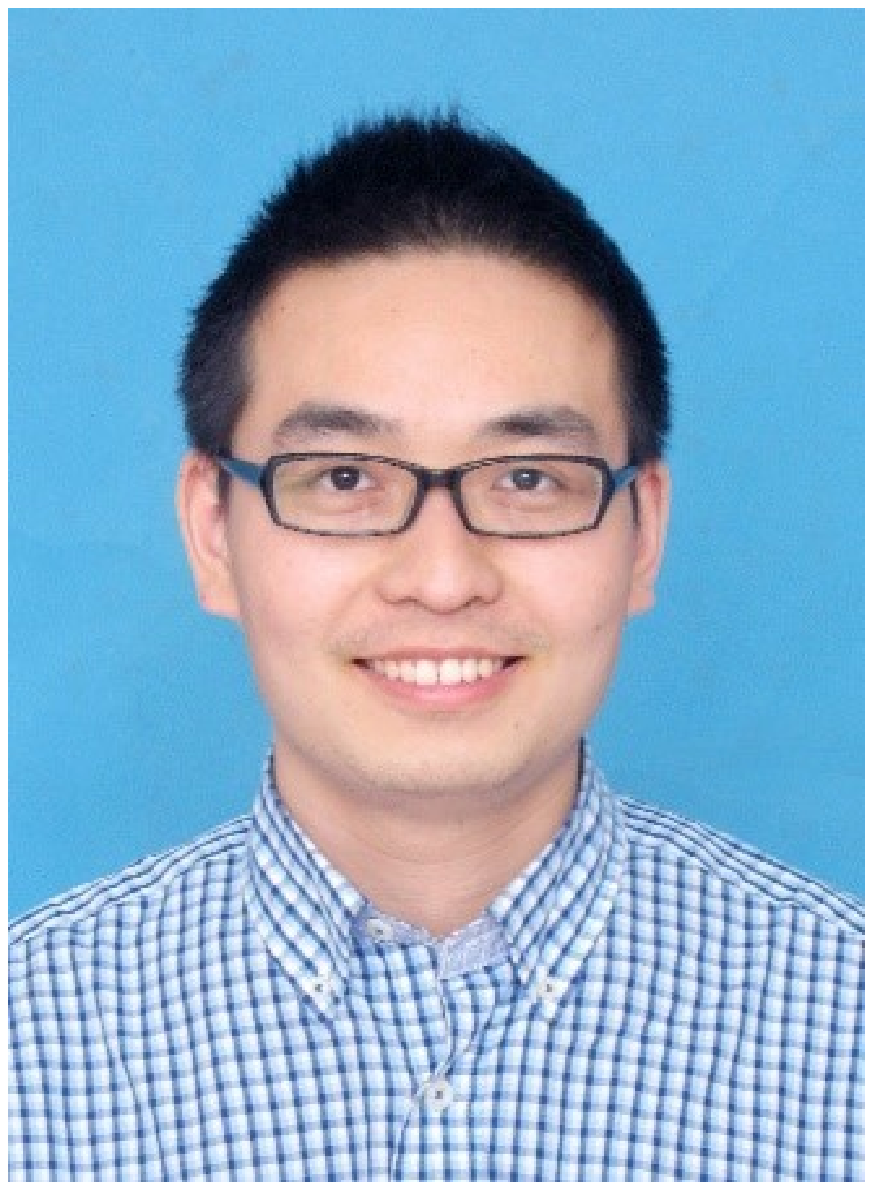}}] {Dingcheng Yang}
received the B.S. degree in electronic engineering
and his Ph.D. degree in Space Physics from Wuhan University, Wuhan, China, in 2006 and 2012.
Now, he is an Associate Professor in the Information Engineering School of
Nanchang University, Nanchang, China. His research interests are cooperation communications, IoT/cyber-physical
systems, UAV communications and wireless resource management. He had published
more than 50 papers including journal papers on IEEE transactions on Vehicular
Technology etc. and conference papers such as IEEE GlOBECOM.
\end{IEEEbiography}
\vspace{-2cm}
\begin{IEEEbiography}[{\includegraphics[width=1in,height=1.25in,clip,keepaspectratio]{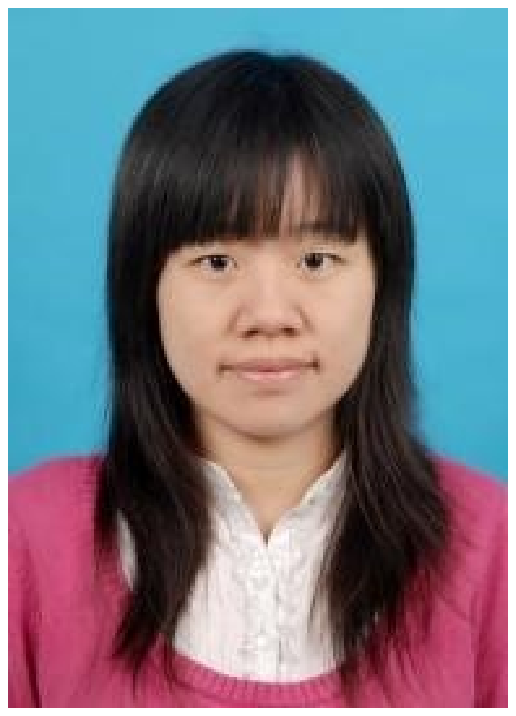}}]
{Lin Xiao} received her Ph.D. degree in the School of Electronic Engineering and Computer Science from Queen Mary University of London in 2010. After that, she worked in China Academy of Telecommunication Research of MITT for one year. Now, she is a Professor in the Information Engineering School of Nanchang University. Her research interests include wireless communication and networks, in particular, UAV network planning and optimization, radio resource management, relay and cooperation communication.
\end{IEEEbiography}
\end{document}